\documentclass[12pt]{article}
\pdfoutput=1
\usepackage{amsmath}
\usepackage{amssymb}
\usepackage{graphicx}
\usepackage{subcaption}
\usepackage{url}
\usepackage{comment}
\usepackage{slashed}

\usepackage[sort&compress, numbers, merge]{natbib}

\usepackage[colorlinks=true, linkcolor=blue, citecolor=blue,
urlcolor=black]{hyperref}

\title{Coannihilation}
                 
\begin{document}
\baselineskip 0.6cm

\def\simgt{\mathrel{\lower2.5pt\vbox{\lineskip=0pt\baselineskip=0pt
           \hbox{$>$}\hbox{$\sim$}}}}
\def\simlt{\mathrel{\lower2.5pt\vbox{\lineskip=0pt\baselineskip=0pt
           \hbox{$<$}\hbox{$\sim$}}}}
\def\simprop{\mathrel{\lower3.0pt\vbox{\lineskip=1.0pt\baselineskip=0pt
             \hbox{$\propto$}\hbox{$\sim$}}}}
\def\tr{\mathop{\rm tr}}
\def\SU{\mathop{\rm SU}}

\begin{titlepage}

\begin{flushright}
IPMU18-0201
\end{flushright}

\vskip 1.1cm

\begin{center}

{\Large \bf 
How Heavy can Neutralino Dark Matter be?
}

\vskip 1.2cm
Hajime Fukuda$^{1}$,
Feng Luo$^{2}$
and 
Satoshi Shirai$^{1}$
\vskip 0.5cm

{\it
$^1${Kavli Institute for the Physics and Mathematics of the Universe (WPI), \\The University of Tokyo Institutes for Advanced Study, The University of Tokyo, Kashiwa
 277-8583, Japan}\\
 $^2${School of Physics and Astronomy, Sun Yat-sen University, Zhuhai 519082, China}
}

\vskip 1.0cm

\abstract{
What is the upper limit of the mass of the neutralino dark matter whose thermal relic is consistent with the observation?
If the neutralino dark matter and colored sparticles are extremely degenerated in mass, with a mass difference less than the QCD scale, the dark matter annihilation is significantly increased and enjoys the ``second freeze-out'' after the QCD phase transition.
In this case, the neutralino dark matter with a mass much greater than 100 TeV can realize the correct dark matter abundance.
We study the dark matter abundance and its detection in the case of such highly degenerated mass spectrum of the neutralino dark matter and colored supersymmetric particles.
}

\end{center}
\end{titlepage}

%%%%%%%%%%%%%%%%%%%%%%%%%%%%%%%%%%%%%%%%%%%%%%%%%%%%
\section{Introduction}
\label{sec:intro}
%%%%%%%%%%%%%%%%%%%%%%%%%%%%%%%%%%%%%%%%%%%%%%%%%%%%
The supersymmetric (SUSY) standard model (SSM) is one of the most attractive candidate of the physics beyond the standard model.
In the SSM, the lightest SUSY particle (LSP) is stable if the $R$-parity is conserved.
The most important feature of the SSM is that the LSP neutralino can be a weakly interacting massive particle (WIMP) dark matter (DM).

The WIMP dark matter abundance is mainly determined by the freeze-out mechanism, if the reheating temperature of the Universe is high enough and there is no additional entropy production such as a moduli field decay.
An interesting feature of the freeze-out mechanism is that the DM abundance can be determined by the DM annihilation rate at the early Universe and predicted solely from the low-energy property of the DM sector, regardless of the initial condition of the Universe.
Generally speaking, as the DM mass gets larger, the DM abundance also gets larger.
In order to keep the thermal relic DM density being consistent with the observation, we can obtain an upper limit of the DM mass.
This upper limit is a very important guideline for the DM search experiments.

In the case of the SSM, the upper limit of the LSP DM mass depends on the details of the mass spectrum.
If the lightest neutralino is the wino, the upper limit of the LSP mass is around 3 TeV \cite{Hisano:2006nn}.
In fact, this wino LSP is a generic prediction of the anomaly mediation model \cite{Randall:1998uk, Giudice:1998xp}.
This model is the simplest realization of the (mini-)split SUSY \cite{Wells:2003tf, *Wells:2004di, ArkaniHamed:2004fb, *Giudice:2004tc, *ArkaniHamed:2004yi, *ArkaniHamed:2005yv},  
which is getting more and more attention \cite{Hall:2011jd, *Hall:2012zp, *Nomura:2014asa, Ibe:2011aa, *Ibe:2012hu, Arvanitaki:2012ps, ArkaniHamed:2012gw}.
The wino has rich signatures of (in)direct dark matter \cite{Hisano:2003ec, *Hisano:2004ds, *Hisano:2005ec, Hisano:2010fy, *Hisano:2010ct, *Hisano:2012wm, *Hisano:2015rsa} and collider searches \cite{Ibe:2006de, *Buckley:2009kv, Asai:2007sw, *Asai:2008sk, *Asai:2008im}.
The detection of the 3 TeV wino will be within reach within the next decade.

In the case of a Higgsino dark matter, the upper limit is around 1 TeV.
The Higgsino DM is also intensively studied.
Future hadron collider and direct detection experiments will enable almost complementary searches up to the 1 TeV Higgsino \cite{Mahbubani:2017gjh,Nagata:2014wma,*Fukuda:2017jmk, Matsumoto:2017vfu,*Matsumoto:2018ioi,Chigusa:2018vxz,DiLuzio:2018jwd}.

If the SSM  mass spectrum is fine-tuned, the relic abundance of the LSP can be exceptionally smaller, accordingly, the upper limit of the DM mass is also increased \cite{Griest:1990kh}.
An example is the so-called Higgs-funnel region, where the neutralino/chargino mass is close to half of the heavier Higgs mass.
The annihilation process of the neutralino is significantly enhanced as the $S$-channel process hits the pole of the heavy Higgs propagator.
In this case the upper limit of the DM mass can be around 10 TeV \cite{Edsjo:1997bg, Profumo:2005xd, Gilmore:2007aq,resInprep}.

Another exceptional case is the coannihilation.
If the lightest neutralino and next-to-LSP (NLSP) is almost degenerated in mass, the effective annihilation rate of the neutralino LSP can be enhanced through the NLSP annihilation.
A prime example is coannihilation with colored sparticles.
In particular the coannihilation with gluino drastically increases the effective annihilation rate and the upper limit of the DM is around 10 TeV \cite{Harigaya:2014dwa, Ellis:2015vaa,Ellis:2015vna,Liew:2016hqo, Nagata:2015hha,*Nagata:2017gci}.
This 10 TeV gluino can be also probed at a future hadron collider \cite{Ellis:2015xba}.

In this paper, we further investigate the coannihilation with colored sparticles.
If the mass difference between the colored NLSP and DM is less than around the QCD scale, the annihilation rate of the DM is affected by non-perturbative QCD effects.
In this case, in addition to the conventional freeze-out which occurs at temperature around $m_{\rm DM}/30$, there is another annihilation stage in the QCD phase transition era.
Due to the non-perturbative strong interaction, the effective annihilation cross section drastically increases, allowing even PeV scale dark matter.
This large mass is above the upper-bound on the WIMP mass $\sim 100$ TeV, which is a generic upper-limit of the WIMP DM mass based on $S$-wave unitarity bound \cite{Griest:1989wd}.
This DM reduction mechanism is studied in the context of the non-standard-model strong interaction \cite{Harigaya:2016nlg} and suggested in the context of the SSM \cite{Profumo:2005xd}.
We revisit this possibility, paying attention to the non-perturbative annihilation and chemical equilibrium between the LSP and colored sparticles.
We also study the detection of such a heavy DM.
Due to the small mass difference with the colored NLSP, the direct detection rate is significantly increased.

\section{Coannihilation}
In this section, we discuss the coannihilation in the case that the colored NLSP is highly degenerated with the neutralino LSP dark matter in mass.
First, we discuss the annihilation processes of the colored NLSP in the high temperature and QCD phase transition eras.
Then, we study the chemical equilibrium condition of the neutralino LSP and the colored NLSP.
In this paper, we mainly focus on the squark-neutralino coannihilation.

\subsection{Squark Annihilation}
\label{seq:ann}
If the mass difference between the neutralino LSP and squark NLSP is less than $\sim 1\,\text{GeV}$, the cosmological evolution of the dark matter shows interesting features.
Once the temperature gets lower than around $m_{\rm DM}/30$, the dark matter coannihilation process freezes out, as in the case of conventional WIMP freeze-out scenario.
However, if the colored sparticle is still abundant in the QCD phase transition era, non-perturbative QCD effect restarts the coannihilation.
When the temperature gets low enough, the coannihilation process freezes out again.
In the following, we describe these two perturbative and non-perturbative freeze-outs\,\footnote{To be precise, the word ``perturbative'' is not very appropriate here, since we include the Sommerfeld and the bound-state effects, which are inherently non-perturbative and important before the QCD phase transition, in the ``perturbative annihilation''. However, in order to be concise when referring to the two phases before and after the QCD phase transition, as well as to highlight the key differences of them, we choose to still use the word ``perturbative annihilation'' to refer to the annihilation process through the QCD in the Coulomb phase and ``non-perturbative annihilation'' to the QCD in the confined phase.}.

\subsubsection{Perturbative Annihilation}
\label{sec:pertann}
We first review the squark-neutralino coannihilation process in the era long before the QCD phase transition. First of all, coannihilation can happen only if the interconversion rate between a squark and a neutralino (including decays/inverse decays, and conversions by scattering with the standard model particles in the thermal bath) is sufficiently large compared to the Hubble expansion rate, otherwise the two particle species will freeze out independently. As we will discuss in section~\ref{seq:che}, this condition of coannihilation can be satisfied in this era, so that to a very good approximation the number density ratio of squarks and neutralinos equals to their thermal equilibrium number density ratio at the same temperature, that is, $n_{\tilde{q}} / n_{\tilde \chi^0_1} = n_{\tilde{q}}^{eq} / n_{\tilde \chi^0_1}^{eq}$. 
Therefore we can use a single Boltzmann equation to track the evolution of the total {\it yield}, which is the number density sum of squarks, antisquarks and neutralinos divided by the entropy density~\footnote{We assume that there is no asymmetry between $\tilde{q}$ and $\tilde{q}^*$, and we take $n_{\tilde{q}} = n_{\tilde{q}^*}$ in our calculation.}, $\tilde{Y}\equiv (n_{\tilde{q}} + n_{\tilde{q}^*} + n_{\tilde \chi^0_1}) / s$, 
\begin{eqnarray}
\frac{d\tilde{Y}}{dx}=-\frac{xs}{H(m_{\tilde \chi^0_1})}\left(1+\frac{T}{3g_{\ast s}}\frac{dg_{\ast s}}{dT}\right)\langle \sigma_{eff} v\rangle\left(\tilde{Y}^2-\tilde{Y}^2_{eq}\right),
\label{eq:bol}
\end{eqnarray}
where
\begin{eqnarray}
x \equiv {m_{\tilde \chi^0_1} \over T}, \, s = {2 \pi^2 \over 45} g_{\ast s} T^3, \; H(m_{\tilde \chi^0_1}) = \left({4 \pi^3 G_N g_\ast \over 45}\right)^{1 \over 2} m_{\tilde \chi^0_1}^2 \, . 
\label{eq:defxsh}
\end{eqnarray}
$G_N$ is the gravitational constant, $m_{\tilde \chi^0_1}$ is the neutralino LSP mass, $g_{\ast s}$ and $g_\ast$ are the numbers of effectively massless degrees of freedom associated with the entropy density and the energy density, respectively. 

In principle, the thermally-averaged effective annihilation cross section, $\langle \sigma_{eff} v\rangle$,  should include contributions of all (co)annihilation channels which can change $\tilde{Y}$. However, for the case of highly degenerated colored NLSP and neutralino LSP we focus in this paper, the dominant contributions to $\langle \sigma_{eff} v\rangle$ come from $\tilde{q} \tilde{q}^*$ annihilations. Also, since we are not committed to the details of the SSM mass spectrum which affects the size of electroweak (co)annihilation channels through for example squark left-right mixing angle, we include only the pure strong interaction channels, $\tilde{q} \tilde{q}^* \to g g$ and $\tilde{q} \tilde{q}^* \to q \bar{q}$, in calculating $\langle \sigma_{eff} v\rangle$. These two processes are universal for all squark flavors and are usually the dominant ones.~\footnote{Ref.~\cite{Ellis:2018jyl} discussed the case when the dominant annihilation channels are $\tilde{t} \tilde{t}^* \to hh, W^+ W^-, ZZ$, when the interactions between stop and Goldstone bosons are large.} We also include in $\langle \sigma_{eff} v\rangle$ the $\tilde{q} \tilde{q}^*$ Sommerfeld and bound-state effects as discussed in detail in Ref.\,\cite{Liew:2016hqo}\footnote{The Sommerfeld and/or bound-state effects are also recently discussed in \cite{deSimone:2014pda, *Ellis:2014ipa, *vonHarling:2014kha, *Ibarra:2015nca, *Cirelli:2016rnw, *Kim:2016zyy, *Kim:2016kxt, *ElHedri:2016onc, *Petraki:2016cnz, *ElHedri:2017nny, *Pierce:2017suq, *Keung:2017kot, *Garny:2017rxs, *Harz:2017dlj, *Biondini:2018pwp, *ElHedri:2018atj, *Binder:2018znk, *Biondini:2018xor, *Biondini:2018ovz, *Harz:2019rro, *Mitridate:2017izz,  Harz:2018csl}.}. With these considerations, we use 
\begin{eqnarray}
\langle \sigma_{eff} v\rangle=\langle \sigma v\rangle_{\tilde{q} \tilde{q}^*} \frac{g_{\tilde{q} \tilde{q}^*}^2(1+\Delta)^3e^{-2\Delta x}}{g_{eff}^2} \, ,
\label{eq:sigmaeff}
\end{eqnarray}
where 
\begin{equation}
\Delta\equiv (m_{\tilde{q}} -m_{\tilde \chi^0_1})/m_{\tilde \chi^0_1} \, , \;\; g_{eff}\equiv g_{\tilde \chi^0_1}+g_{\tilde{q} \tilde{q}^*}(1+\Delta)^{3/2}e^{-\Delta x} \, ,
\end{equation} 
where $g_{\tilde \chi^0_1} = 2$ is the degrees of freedom for a neutralino, assumed to be a bino. $g_{\tilde{q} \tilde{q}^*} = 6$ is the sum of the degrees of freedom for squark and antisquark. $\langle \sigma v\rangle_{\tilde{q} \tilde{q}^*}$ is the thermally-averaged annihilation cross section times relative velocity~\footnote{As is noted in \cite{Liew:2016hqo}, since $g_{\tilde{q} \tilde{q}^*} = 6$ is the sum of the degrees of freedom for squark and antisquark, a factor of $1/2$ needs to be put in $\langle \sigma v\rangle_{\tilde{q} \tilde{q}^*}$, if using the usual spin and color averaged cross section. This is explained in the Appendix of \cite{Srednicki:1988ce}.}, given as
\begin{equation}
\langle \sigma v\rangle_{\tilde{q} \tilde{q}^*} = \langle \sigma v (\tilde{q} \tilde{q}^* \to gg, q\bar{q})\rangle + \langle \sigma v \rangle_{bsf}\frac{\langle\Gamma \rangle_\eta}{\langle\Gamma \rangle_\eta+\langle\Gamma \rangle_{dis}} \, .
\label{eq:xxtosm}
\end{equation}
In the above expression, the first term is the sum of the Sommerfeld enhanced $\tilde{q} \tilde{q}^* \to gg$ and $\tilde{q} \tilde{q}^* \to q\bar{q}$ cross sections, and the second term is the contribution from the $\tilde{q} \tilde{q}^*$ bound-state effect~\footnote{We update in this work the bound-state formation cross section and dissociation rate calculations compared to \cite{Liew:2016hqo} by including the contribution from the diagram of gluon emission from gluons. This type of diagram is only present when the gauge interaction is non-abelian, and was first pointed out in~\cite{Asadi:2016ybp}. We follow~\cite{Harz:2018csl} for the calculation.}, where $\langle \sigma v \rangle_{bsf}$ is the bound state formation cross section, $\langle\Gamma \rangle_\eta$ is the bound state annihilation decay rate, and $\langle\Gamma \rangle_{dis}$ is the bound state dissociation rate. All quantities are thermally averaged. We note that the bound-state effect we are discussing in this subsection is in the perturbative regime of QCD at high  temperature, in contrast to the QCD bound state which we will discuss in the next subsection, when the QCD enters the non-perturbative regime after the QCD phase transition. For the detail calculations in the perturbative regime, we refer the readers to Ref.~\cite{Liew:2016hqo, Harz:2018csl}.

We evaluate Eq.~(\ref{eq:bol}) from $T \sim m_{\tilde \chi^0_1}$ to lower temperatures. The usual freeze-out (when $\tilde{Y}/\tilde{Y}_{eq} - 1$ becomes larger than 1) happens around $T \sim m_{\tilde \chi^0_1} / 30$. After that, $\tilde{Y}$ continues decreasing, and in particular the bound-state effect becomes effective when $T$ is around and below the bound-state binding energy, $E_B \sim m_{\tilde \chi^0_1} / 100$. $\tilde{Y}$ essentially ceases to decrease after $T \sim m_{\tilde \chi^0_1} / 1000$, until the new annihilation process kicks in after the QCD phase transition which we describe in the next subsection.  

\subsubsection{Non-perturbative Annihilation}
\label{sec:nonpertann}

\begin{figure}[h]
\centering
\includegraphics[clip, width = 0.4 \textwidth]{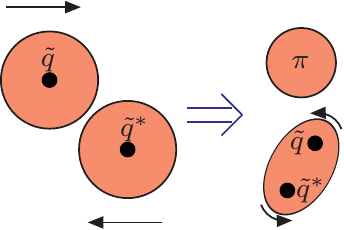} 
\caption{Schematic picture of non-perturbative annihilation.}
\label{fig:brwon}
\end{figure}

After the first freeze-out, as the Universe gets cooler, the QCD phase transition occurs.
With non-perturbative QCD effect, the squark annihilation rate drastically increases.
In this subsection, we review the annihilation process using the non-perturbative effect of the QCD based on Ref.\,\cite{Kang:2006yd}\,\footnote{Also, see Ref.\,\cite{Geller:2018biy} for the analytic calculation, which is basically consistent with Ref.\,\cite{Kang:2006yd}.}.
The annihilation process is in the following way. First, a squark forms a QCD bound state with a quark/gluon, which are abundant in the environment, and becomes a SUSY hadron such as a mesino.
The SUSY hadron has a large geometrical size $\Lambda_{\rm QCD}^{-1}$ and large scattering cross section with another SUSY hadron.
Once such a scattering occurs, a squark-squark bound state ${\tilde{q}}^*\tilde{q}$ of a large angular momentum forms. Then, the bound state de-excites into the ground state. Finally, the $S$-wave state annihilates into quarks and gluons. 
If the rates of de-excitation and annihilation of the ${\tilde{q}}^*\tilde{q}$ bound state are fast enough, the annihilation cross section of the squarks is virtually equivalent to the bound state formation cross section, which is much larger than the perturbative annihilation.
In Fig.~\ref{fig:brwon}, we show a schematic picture of this annihilation process.
In the following, we examine the rates of these non-perturbative processes.

First, we discuss the ${\tilde{q}}^*\tilde{q}$ bound state formation. Soon after the QCD phase transition at temperature $T_c \simeq 200\,\text{MeV}$, the squarks start to form mesinos, for example, $\tilde{q} \bar{q}$ and ${\tilde{q}}^* {q}$. 
Since the squark is much heavier than the QCD scale, the squark is localized at the center in the mesino bound state, $\sim m_{\tilde q}^{-1}$ in radius, and the quark/gluon partons surround it with a radius of $R_\text{had} = \Lambda_\text{had}^{-1}$, where $\Lambda_\text{had}$ is the hadronic mass scale.
Then, these heavy hadrons interact with each other, forming a squarkonium ${\tilde{q}}^*\tilde{q}$, a squark-antisquark bound state, where the cross section is the geometrical one,
\footnote{Note that Ref.\,\cite{Jacoby:2007nw} has claimed that $\tilde{q}qq$ bound states are involved in the squarkonium formation process instead of an anti-mesino, $\tilde{q}\bar{q}$. However, they concluded that even including $\tilde{q}qq$ state, the squarkonium formation rate is not so different. Thus, we simply use Eq.\,(\ref{eq:form}) here. }
\begin{eqnarray}
\label{eq:form}
\sigma_\text{form} ( (\tilde{q} \bar{q}) + ({\tilde{q}}^* {q}) \to  ({\tilde{q}}^*\tilde{q}) + X  ) \sim \pi \Lambda_\text{had}^{-2}.
\end{eqnarray}
This geometrical cross section can be explained in another way. As long as the impact parameter of the collision is less than the size of the mesino, $R_\text{had}$, we may assume the interaction is strong enough. Thus, the cross section for a partial wave $L$ saturates the unitarity bound if $L < L_\text{max} \equiv p R_\text{had}$, where $p$ is the squark momentum. Summing up the partial wave cross section up to $L_\text{max}$, we obtain the geometrical cross section\,\cite{SakuraiQM};
\begin{align*}
    \sigma_\text{form} \sim \sum_{L}^{L_\text{max}} \frac{4\pi (2 L + 1)}{p^2} \simeq \pi\frac{L_\text{max}^2}{p^2} = \pi R_\text{had}^2.
\end{align*}
Ref.\,\cite{Geller:2018biy} also discusses this behavior analytically.

Then let us estimate $R_\text{had} = \Lambda_\text{had}^{-1}$. In Ref.\,\cite{Kang:2006yd}, $\Lambda_\text{had}$ is taken to be $1\,\text{GeV}$. On the other hand, the pion charge radius is experimentally $0.67\,\text{fm} \sim (300\,\text{MeV})^{-1}$\,\cite{Tanabashi:2018oca}, which suggests to use $\Lambda_\text{had} \sim 300\,\text{MeV}$. Also, Ref.\,\cite{Jacoby:2007nw} uses a slightly larger value, $R_\text{had} \sim 0.5 \,\text{fm} \sim (400\,\text{MeV})^{-1}$. Moreover, Ref.\,\cite{Kusakabe:2011hk} also claims that such larger value can be available\,\footnote{They also discuss the energy dependence of the cross section. However, in our region of interest, the annihilation process is almost instantaneous in the temperature and we do not discuss it.}. Thus, combining these contexts, we take the $\Lambda_{\rm had}$ as a free parameter  which varies from $400$ MeV to $1$ GeV in the following analysis.

Next, we discuss the de-excitation process. Initially, the bound state has a very large angular momentum: since the  momentum of the squark is $p \sim \sqrt{2 m_{\tilde{q}} T}$, and the impact parameter is $r \sim \Lambda_\text{had}^{-1}$, the angular momentum is typically
\begin{eqnarray}
L \sim 10  \left(\frac{m_{\tilde{q}}}{\text{TeV}}\right)^{\frac12} \left(\frac{T}{T_c}\right)^{\frac12} \left(\frac{\Lambda_{\rm had}}{1~{\rm GeV}}\right)^{-1}.
\end{eqnarray}
The formed bound state is a rather excited state with a large angular momentum $L$.
In order that the bound state annihilates efficiently, it must de-excite into the lower $L$ state.
Typically, after a bound state loses $\mathcal{O}(L)$ angular momenta and falls into a deep bound state, the energy gaps between the excited states are larger than the hadronic scale and the bound state can immediately de-excite through spontaneous pion emission. Thus, we below discuss the de-excitation process for the large $L$ bound state to lose $\mathcal{O}(L)$ angular momenta. In terms of the energy, the large $L$ bound state loses $\mathcal O(\Lambda_\text{had})$.

Ref.\,\cite{Kang:2006yd} estimates the de-excitation rate for the squark by the dipole emission of photons as
\begin{eqnarray}
\label{eq:deext}
\Gamma_\text{de-excitation} \sim \frac{Q^2 \alpha \Lambda_\text{had}^3}{\alpha_\text{QCD}(\Lambda_\text{had}) m_{\tilde q}^2},
\end{eqnarray}
where $Q$ is the electromagnetic charge of the squark, $\alpha$ is the electromagnetic structure constant and $\alpha_\text{QCD}(\mu)$ is the QCD structure constant at an energy scale $\mu$.
This formula comes from the classical electrodynamics. Since the angular momentum and the quantum number of the bound state is very large, it is reasonable to regard them as a classical bound state.
Ref.\,\cite{Kusakabe:2011hk} also estimates a similar formula for the radiative de-excitation\,\footnote{Although they do not use higher angular momentum region, as Ref.\,\cite{Kang:2006yd} has discussed, it takes longer for higher angular momentum state to de-excite and that process is more important here.}.

Also, Refs.\,\cite{Jacoby:2007nw} and \cite{Kusakabe:2011hk} point out the existence of collisional de-excitation process, where electromagnetic or hadronic collision cools the squarkonium down,
\begin{eqnarray}
({\tilde{q}}^*\tilde{q}) + X \to ({\tilde{q}}^*\tilde{q})' + X.
\end{eqnarray}
Here, $X$ is $\gamma$ or hadrons.
It is expected that these cross sections roughly are equal to the geometrical cross section,
\begin{eqnarray}
\label{eq:deext_col}
\sigma_\text{de-excitation, col, $\pi$} &\sim& 
\pi R_\text{had}^2 ,\\ 
\sigma_\text{de-excitation, col, $\gamma$} &\sim&  \alpha^2 \pi R_\text{had}^2.
\end{eqnarray}
This cross section is estimated just like the formation cross section\,\footnote{Unlike a model discussed in Ref.\,\cite{Geller:2018biy}, the quark cloud has the electromagnetic charge and the electromagnetic collision is suppressed only by $\alpha$.}.
Note that Refs.\,\cite{Jacoby:2007nw} and \cite{Kusakabe:2011hk} discuss that the cross section should be smaller than the destruction cross section by a factor of $\left({\alpha_{\rm QCD}(\Lambda_\text{had}) E_X/\Lambda_{\rm had}}\right)^2$, where $E_X$ is the energy of $X$. On the other hand, if we apply the naive dimensional analysis\,\cite{Manohar:1983md}, these rates may be comparable to each other.
This collisional cooling needs $O(L)$ collisions to loose angular momentum enough.
In the following we adopt the estimation of Ref.\,\cite{Jacoby:2007nw}.

Just as the de-excitation process, particles in the thermal bath may excite the bound state or even break the bound state. For the transition between the bound states, if the de-excitation rate, Eq.\,(\ref{eq:deext}), is faster than the cosmic expansion, the bound states are at least in the chemical equilibrium. Since the energy gap between bound states with lower angular momenta is of order $O(E_B) \gg T_c$, where $E_B$ is the typical binding energy for the lower angular momentum bound states and satisfies $E_B \sim \alpha_\text{QCD}^2(E_B) m_{\tilde{q}}$, and this corresponds to the gap of chemical potentials for the bound states, we conclude that the number of the $S$-wave bound state is exponentially larger than the others. Thus, as in Ref.\,\cite{Kang:2006yd}, we do not need to include the effect as long as the de-excitation is effective enough.

On the other hand, the destruction of the bound state into SUSY particles should be also considered, as these processes cannot reduce the number of the SUSY particles.
If the decay/annihilation rate of the bound state is much greater than the Hubble rate, we can treat this resonance as an intermediate state of the squark annihilation.
Then the effective annihilation cross section $\sigma_\text{form}'$ can be written as
\begin{eqnarray}
\sigma_\text{form}'(T) =  {\rm BF}( (\tilde q^* \tilde q) \to {\rm SM's}) \times \sigma_\text{form},
\end{eqnarray}
and the $ {\rm BF}( (\tilde q^* \tilde q) \to {\rm SM's})$ is the effective branching fraction of the squarkonium decays into SM particles and written as
\begin{align}
{\rm BF}( (\tilde q^* \tilde q) \to {\rm SM's} )= \frac{\langle \Gamma_\text{de-excitation} \rangle }{\langle \Gamma_\text{de-excitation} \rangle+ \langle \Gamma_\text{destruction}\rangle},
\label{eq:BF}
\end{align}
where $\Gamma_\text{destruction}$ is the destruction rate of $\tilde{q}^*\tilde{q}$ bound state and $\langle \cdots \rangle$ represents thermal average. The destruction rate is the sum of the decay rate of the squark, $\Gamma_\text{decay}$, and the collisional process, $\Gamma_\text{dest, coll}$,
\begin{eqnarray}
\label{eq:dest}
\Gamma_\text{destruction} = \Gamma_\text{decay} + \Gamma_\text{dest, coll}.
\end{eqnarray}
The constituent squark in the bound state can decay into the LSP, if the binding energy is smaller than the mass difference between the squark and LSP. 
In this estimation, we adopt
\begin{align}
\Gamma_{\rm decay} = 2 \times \Gamma( \tilde q  \to  \tilde \chi^0_1 +{\rm SM}).
\end{align}
The factor $2$ is for the decay of both squark and anti-squark.
In our estimation, in order to take the binding energy into account, we estimate the decay width of the squark mesino by replacing the mass difference, $\Delta m \equiv m_{\tilde{q}} - m_{\tilde \chi^0_1}$, with $\Delta m - \Lambda_{\rm had}$.
This replacement can be understood as follows; first, consider the non-relativistic effective theory of the squarkonium bound state, just like Ref.\,\cite{Hisano:2003ec}. Then, the width of the bound state comes from the imaginary part of the squark propagator. Only the difference from the squark decay width is the momentum squared, which now corresponds to replacing the mass diffrence as $\Delta m - \Lambda_{\rm had}$. Also, for the beta decay of various atomic nuclei, a similar calculation of the width is known to be consistent with the observation\,\cite{Wong:1998ex}.

For the collisional process, the dominant one for $T\gtrsim O(10)\,\text{MeV}$ is the inverse process of the bound state formation $ \pi + (\tilde q^* \tilde q) \to (\tilde q^* q) + (\tilde q \bar{q}) $ although it is suppressed by the Boltzmann factor compared with the formation rate, since we need additional binding energy around $\Lambda_\text{had}$ to break the bound state apart into mesinos.
The process is regarded as the collision between finite-size objects, just like the bound state formation, and
the cross section, $\sigma_\text{dest, coll, break}$, is expected to be geometrical: $O(\pi R_{\rm had}^2)$.

We can also derive the result by applying the Milne relation\,\footnote{See Eqs.\,(19)-(21) in Ref.\,\cite{Ellis:2015vaa}, for example.} to Eq.\,\eqref{eq:form}; the Milne relation is
\begin{align*}
\sigma_\text{dest, coll, break} = \frac{p^2}{\omega^2} \frac{g_{\tilde{q}\bar{q}}^2}{g_{\tilde{q}\tilde{q}^\star} g_\pi} \sigma_\text{form},
\end{align*}
where $p$ is the typical momentum of the mesino for the bound state formation, $\omega$ is the typical energy of the pion for the bound state destruction, being around the hadronic scale, and $g_i$ is the degree of freedom for particle $i$. 
As we have discussed above, the bound state formation process can be decomposed into the partial wave and with the maximal angular momentum for the bound state formation $L_\text{max}$, $\sigma_\text{form} \sim \pi L_\text{max}^2 / p^2$. The resultant squarkonium has an angular momentum up to $L_\text{max}$ and thus $g_{\tilde{q}\tilde{q}^\star} \sim \sum_{L}^{L_\text{max}} (2L+1) \sim L_\text{max}^2$.
The cross section is then
\begin{align}
\label{eq:dest_col}
\sigma_\text{dest, coll, break} \sim \pi R_{\rm had}^2,
\end{align}
and indeed geometrical.

However, the relevant number density of the hadron to be multiplied to the cross section
in Eq.\,(\ref{eq:dest_col}) to obtain the reaction rate includes the Boltzmann suppression factor for the meson to be as energetic as $E\gtrsim\Lambda_\text{had}$.
Consequently, the destruction rate can be esimated as
\begin{align}
\Gamma_\text{dest, coll} \simeq \sigma_\text{dest, coll, break} \times n_{\pi,E\gtrsim \Lambda_{\rm had}} \sim \pi R_{\rm had}^2  \Lambda_{\rm had}^2 T  \exp(-\Lambda_{\rm had}/T),
\end{align}
and this destruction rate is suppressed for the temperature much less the hadron scale.

Also, hadronic or electromagnetic collisions can make the conversion of $\tilde q + \bar{q} \to \tilde \chi^0_1$ through which a squarkonium can change into a mesino and a LSP.
This collision rate is roughly proportional to the temperature $T$.
This process may be suppressed if the binding energy is much larger than the squark-DM mass difference.

After the de-excitation, squarks annihilate efficiently. The decay rate is similar to the case of a positronium and given as
\begin{eqnarray}
\Gamma_\text{annihilate} \sim \frac{\alpha_\text{QCD}^2(m_{\tilde q})}{m_{\tilde q}^2} \left|\psi(0)\right|^2,
\end{eqnarray}
where $\psi$ is the wave function for the squarks in the bound state. For the ground state, this reduces to $\Gamma_\text{annihilate} \sim \alpha_\text{QCD}^2(m_{\tilde q})\alpha_\text{QCD}^3(a_{\rm Bohr}^{-1})  m_{\tilde q}$, where $\alpha_\text{QCD}$ from the wave function is evaluated at the scale of inverse Bohr radius of the bound state, $a_{\rm Bohr}^{-1} \sim \alpha_\text{QCD}(a_{\rm Bohr}^{-1}) m_{\tilde q}$. Since this is much larger than the rate to break such deeply bounded state, which is highly suppressed by the Boltzmann factor, $\exp(-E_B / T)$, and the Hubble expansion rate, we can assume that the annihilation is instantaneous once bound states fall into the ground state.

\begin{figure}[htb]
	\centering
	\subcaptionbox{\label{fig:bf_ur}Right-handed scalar up $\tilde u_R$.}{\includegraphics[width=0.47\textwidth]{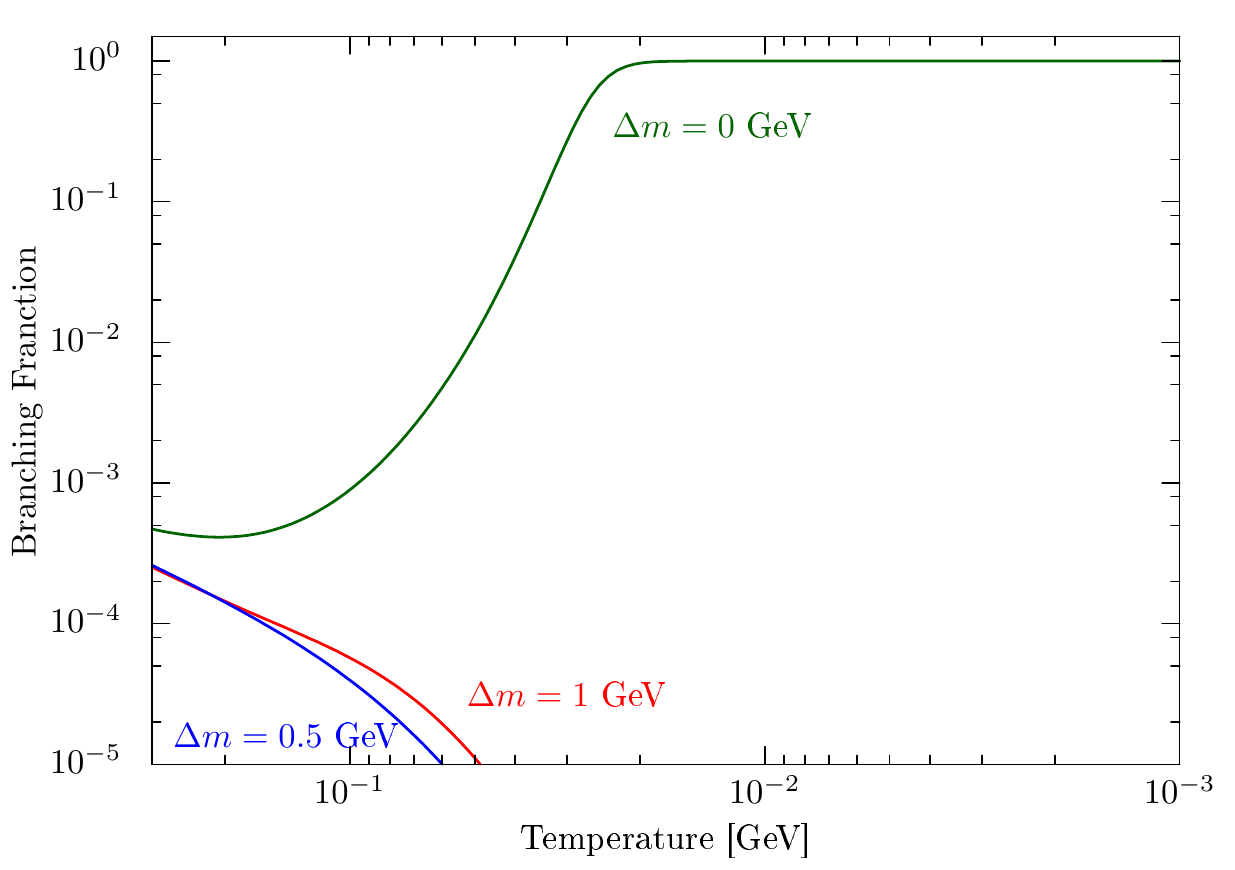}}
	\subcaptionbox{\label{fig:bf_tr}Right-handed scalar top $\tilde t_R$ with flavor violation $\delta_{13} = 10^{-3}$.}{\includegraphics[width=0.47\textwidth]{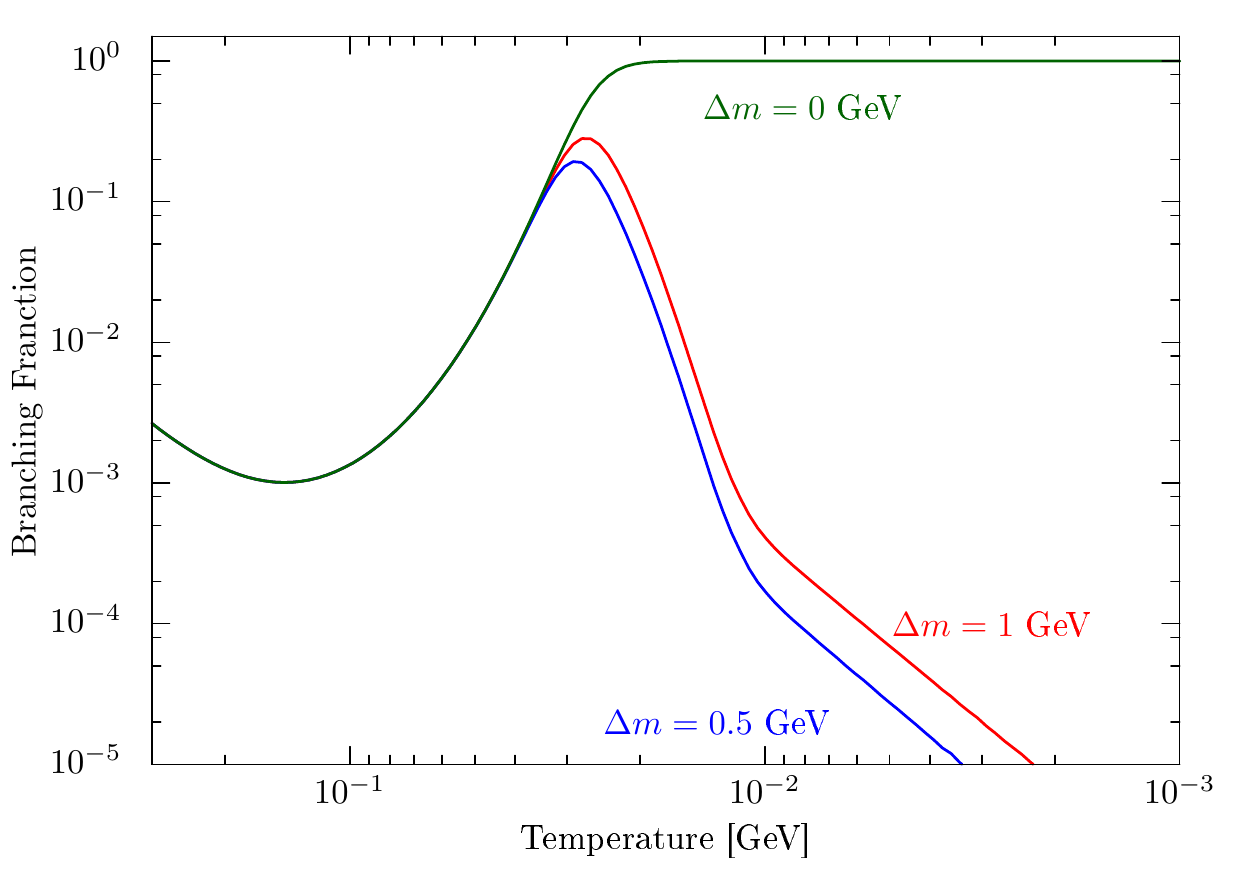}}
\caption{
The effective branching fraction of the $\tilde q^* \tilde q$ bound state decay into SM particles ${\rm BF}( (\tilde q^* \tilde q) \to {\rm SM's} )$.
Here we set $m_{\rm DM} = 10^6$ GeV. 
}
\label{fig:branching_fraction}
\end{figure}

In Fig.~\ref{fig:branching_fraction}, we show the effective branching of the squarkonium \eqref{eq:BF} as a function of the temperature. 
Here we set $\Lambda_{\rm had} = 0.4$ GeV.
In Fig.~\ref{fig:bf_ur}, we show the case of $\tilde u_R$ NLSP with the squark-DM mass difference $\Delta m = 0, 0.5$ and 1 GeV in green, blue and red lines, respectively.
For the large mass difference case, the internal conversion $\tilde q \to \tilde \chi^0_1$ by either the decay or collision dominates for lower temperature region.
In the case of $\Delta m =0$, this internal conversion is prevented due to the squarkonium binding energy.
In Fig.~\ref{fig:bf_tr}, we show the case of $\tilde t_R$ with a flavor mixing with the first generation $\delta_{13} = 10^{-3}$.
With this mixing, the internal conversion is prevented and the effective branching fraction is greater than the $\tilde u_R$ case.

If the NLSP squark is left-handed, we need to consider the mass difference between the up-type and down-type left-handed squarks.
The mass difference comes from the Higgs coupling and the radiative correction. 
For the Higgs coupling, the mass difference is roughly\,\cite{Martin:1997ns}
\begin{eqnarray}
m_{\tilde u_L} - m_{\tilde d_L}  \simeq  \frac{(m_u^2 - m_d^2) + \cos(2\beta) m^2_{W}}{2 m_{\tilde q}}  \sim
100~{\rm MeV} \left(\frac{m_{\tilde q}}{10~{\rm TeV}}\right)^{-1},
\end{eqnarray}
whereas for the radiative correction\,\cite{Cirelli:2005uq}, it is
\begin{eqnarray}
\Delta m \sim 70\,\text{MeV}.
\end{eqnarray}
Thus, if the squark is heavier than $10\,\text{TeV}$, then both up-type and down-type squarks contribute to the annihilation in the QCD era.

Finally, we briefly comment on the squark-squark bound state, $\tilde{q}\tilde{q}q$, for example. Roughly speaking, the formation cross section should be geometrical, but the $Q$-value is different from the  $\tilde{q}^*\tilde{q}$ bound state formation process, and thus whether they are formed or not is not obvious\,\cite{Kusakabe:2011hk}. However, once they are formed, the de-excitation and destruction processes will be similar to the $\tilde{q}^*\tilde{q}$ state case.
As in the case of the  $\tilde{q}^*\tilde{q}$ bound state, if the de-excitation process continues, the final form of the $\tilde{q}\tilde{q}q$ bound state is a hadron with rotating squark cores.
Unlike $\tilde{q}^*\tilde{q}$, because of the wavefunction symmetrization, the angular momentum between squarks must change by $2$ during the de-excitation process, but at least by the collisional process can efficiently de-excite them eventually to the $P$-wave ground state.
Finally, squarks annihilate exchanging virtual neutralino, whose rate is much larger than the Hubble. Hence, we expect that if they are formed, the annihilation process is not much different from the case of the $\tilde{q}^*\tilde{q}$ bound state.

\subsection{Chemical Equilibrium}
\label{seq:che}
In order that the coannihilation processes reduce the LSP abundance significantly, the LSP and squark NLSP must maintain the chemical equilibrium when the squark effective annihilation is efficient.
For higher temperature, the chemical equilibrium is easily maintained.
However, in the QCD phase transition era, the temperature of the Universe is rather low.
We examine the condition of the chemical equilibrium.

\begin{figure}[t]
	\centering
	\subcaptionbox{\label{fig:chemical_strong}QCD interaction}{\includegraphics[width=0.4\textwidth]{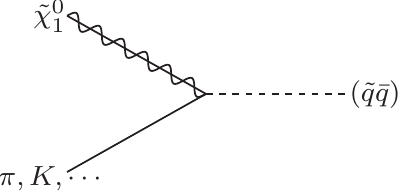}}
	\qquad
	\subcaptionbox{\label{fig:chemical_weak}Weak interaction}{\includegraphics[width=0.4\textwidth]{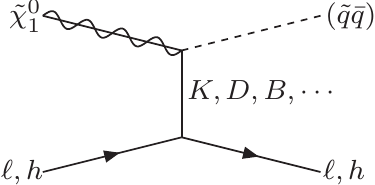}}
\caption{
Diagrams for the conversion between the neutralino and squark.
}
\label{fig:chemical}
\end{figure}

The conversion via  hadron collisions (Fig.~\ref{fig:chemical_strong}) is the most efficient conversion process.
The rate for $\tilde \chi^0_1 \to \tilde q$ is 
\begin{eqnarray}
\label{eq:conv}
\Gamma_\text{conv} \sim c^2 \alpha T   \times 
\begin{cases}
\exp(-m_h/T) & {\rm for~} m_h \gg \Delta m\\
\exp(-\Delta m/T) &  {\rm for~} m_h \ll \Delta m
\end{cases}
\end{eqnarray}
where $m_h$ is the incoming hadron mass and $c$ denotes the hypercharge of the squark. We take the pion mass $m_\pi$ as $m_h$ if the NLSP squark is the first generation, $\tilde{u}$ or $\tilde d$. For the other generations, there is another Boltzmann factor such as $\exp(-m_{B}/T)$, as the initial of final state needs to include heavy flavor mesons, if there is no flavor violation in the squark sector. 
If the mass difference is larger than the meson mass, the rate is  Boltzmann suppressed. We also have the weak process or electromagnetic process, but they are generally smaller than Eq.\,(\ref{eq:conv}), given that the mass difference is the same order as $m_\pi$. For Higgsino or wino LSP, we need to use appropriate gauge/Yukawa couplings instead of $c^2\alpha$.

Charged flavor violation coupling via the weak interaction (Fig.~\ref{fig:chemical_weak}) can be relevant for the heavy flavor scalar quark.
This process is suppressed by the CKM mixing and the Fermi constant.
However there is no significant Boltzmann suppression due to the heavy flavor meson mass, this process can dominate in low temperature region.

\begin{figure}[htb]
	\centering
	\subcaptionbox{\label{fig:conv_ur}Scalar up quark  with $\Delta m=$ 0, 0.5 and 1 GeV.}{\includegraphics[width=0.47\textwidth]{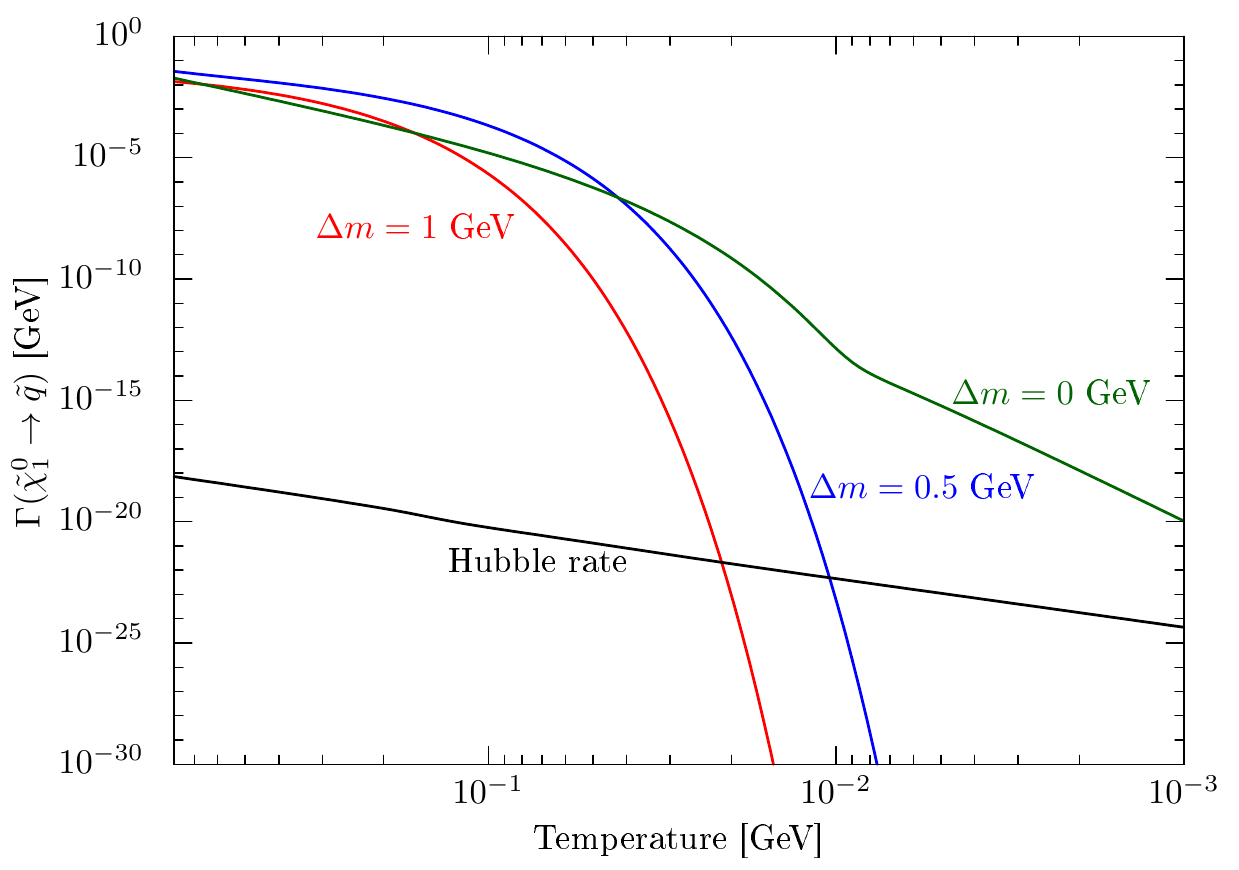}}
	\subcaptionbox{\label{fig:conv_h}Heavy flavor scalar quark with $\Delta m = 0$ GeV.}{\includegraphics[width=0.47\textwidth]{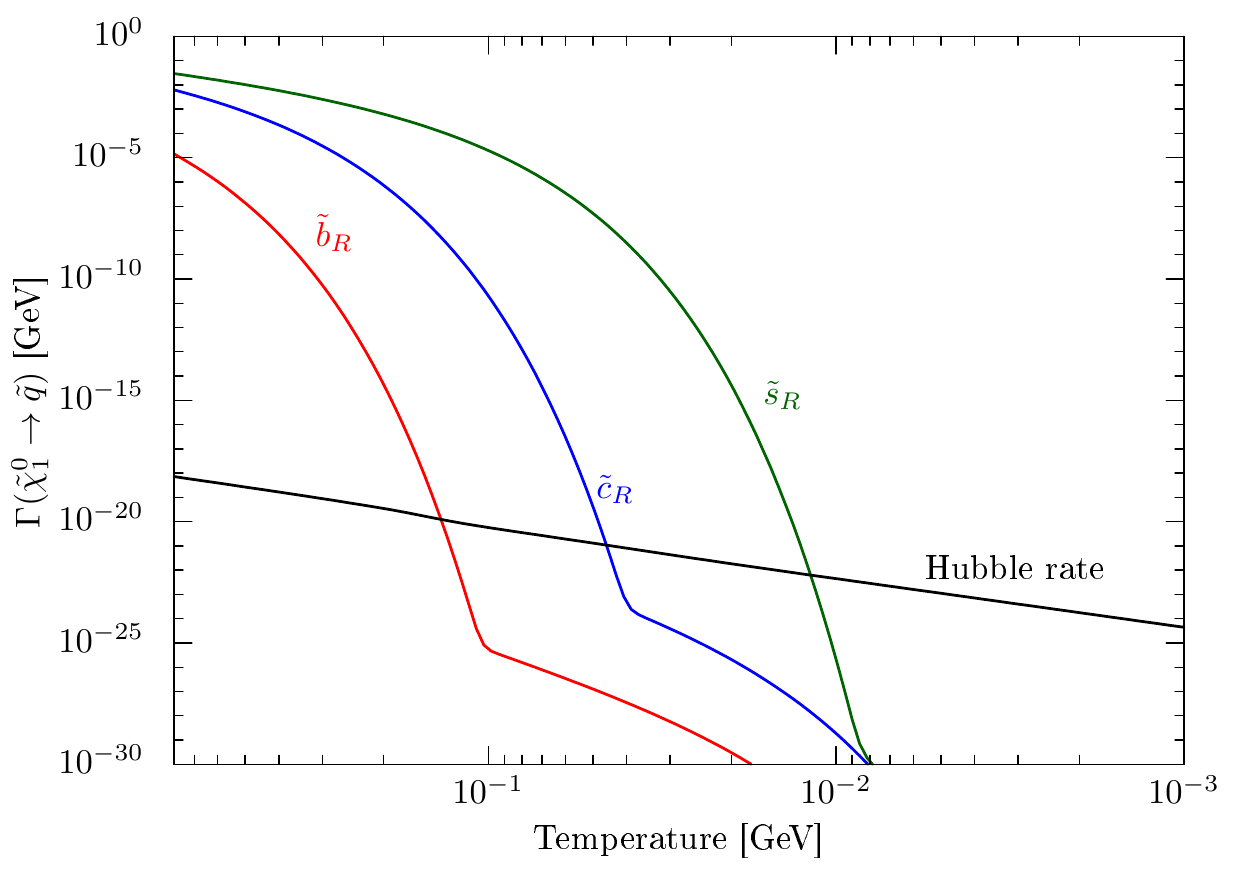}}
\caption{
Conversion rates between the bino LSP and scalar right-handed quark NLSP. 
(a): Scalar up quark.
The green, blue and red lines represent the cases of the mass difference between the NLSP and LSP $\Delta m$ being 0, 0.5, and 1 GeV, respectively.
(b): Heavy flavor scalar quark with $\Delta m=0$ GeV and minimal flavor violation.
The green, blue and red lines represent the cases of $\tilde s_R$, $\tilde c_R$ and $\tilde b_R$, respectively.
}
\label{fig:rate}
\end{figure}

In Fig.~\ref{fig:rate}, we show the conversion rate of $\chi^0_1 \to \tilde q$.
In Fig.~\ref{fig:conv_ur}, we show the cases of right-handed scalar up quark, for the LSP-NLSP mass difference 0, 0.5 and 1 GeV.
For larger mass differences, the conversion rate is suppressed by the Boltzmann factor.
Still, the chemical equilibrium can be maintained even if the temperature is around 10 MeV.
In Fig.~\ref{fig:conv_h}, we show the case of heavy flavor scalar quarks, with minimal flavor violation assumed.
Generally speaking, in the SSM, there can be non-minimal flavor violations.
If there is a flavor mixing between the first generation and third generation with a mixing parameter $\delta_{13}$, the conversion rate of stop is increased as $\Gamma( \tilde \chi^0_1 \to \tilde t_R ) \sim \delta_{13}^2 \Gamma( \tilde \chi^0_1 \to \tilde u_R )$.
Even if the flavor mixing is tiny, the chemical equilibrium can be realized.
For instance, when the mixing parameter $\delta_{13}$ is greater than $O(10^{-6})$, the scalar top can be in chemical equilibrium with the bino LSP for $T\gtrsim 10$ MeV and $\Delta m \lesssim 1$ GeV.

\begin{figure}[h]
\centering
\includegraphics[clip, width = 0.7 \textwidth]{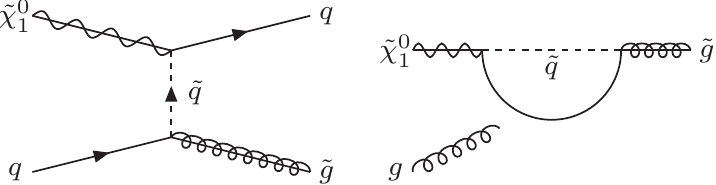} 
\caption{Diagrams for the conversion between the neutralino and gluino.}
\label{fig:gluino_chemical}
\end{figure}

Next, let us comment on the conversion process for the gluino NLSP scenario. If gluino is the NLSP, the virtual squark is involved in the conversion process, as is shown in Fig.~\ref{fig:gluino_chemical}. The conversion rate between the bino LSP and the gluino NLSP is then
\begin{eqnarray}
\Gamma_\text{conv, tree} \sim c^2 \alpha \frac{T^5}{m_{\tilde{q}}^4} \exp(-m_h / T)
\end{eqnarray}
for the tree level process with zero mass difference, where $c$ and $m_{\tilde{q}}$ are the hypercharge and mass of the lightest squark, respectively, and $m_h$ is the mass of hadron involved in the squark-quark-gluino vertex.
Requiring that the conversion rate is greater than the Hubble rate at least around $T\sim T_c$, we obtain
\begin{eqnarray}
m_{\tilde{q}} \lesssim O(1)~\text{TeV}
\end{eqnarray}
for the first generation squark mass. Since this must be heaver than the LSP and the NLSP, we conclude that the gluino must be lighter than TeV for the bino LSP and the gluino NLSP scenario. Also, even if the LSP is wino, the mass bound is small as well.
For such cases, the perturbative annihilation is effective enough to reduce the abundance, as discussed in Sec.\,\ref{sec:pertann}.

On the other hand, if the Higgsino is the LSP and the gluino is the NLSP, the loop-level conversion rate is rather large, as the gluino-Higgsino-gluon dipole operator is relatively enhanced \cite{Toharia:2005gm,Gambino:2005eh,Sato:2012xf,*Sato:2013bta}.
The dominant contribution comes from the stop-top loop and the conversion rate is roughly given by
\begin{eqnarray}
\Gamma_\text{conv, loop} \sim \frac{T^3 m_{t}^2}{(16 \pi^2)^2 m_{\tilde{t}}^4} \exp(-m_\pi / T).
\end{eqnarray}
Then, comparing with the Hubble expansion rate at $T=T_c$, we obtain
\begin{eqnarray}
m_{\tilde{t}} < O(10)\,\text{TeV}.
\end{eqnarray}
Thus, it may be possible to exceed the perturbative annihilation for this case. We will estimate the abundance in future \cite{gluinoInprep}.

\subsection{Relic Abundance}

Now, we calculate the relic abundance, combining the dynamics discussed above. First, we compute the yield for superparticles based on Sec.\,\ref{sec:pertann}. Then, we need to keep track of the non-perturbative annihilation process described in Sec.\,\ref{sec:nonpertann}.

For the non-perturbative annihilation process, we need to solve the Boltzmann equation for coannihilation between the LSP and super-hadrons.
The Boltzmann equations are given as:
\begin{align}
\frac{d n_{\tilde q}}{dt} =&  -3 H n_{\tilde q} - \langle \sigma_{\rm form} v_{\rm rel} \rangle    \left( \frac{1}{2}{\rm BF}( (\tilde q^* \tilde q) \to \tilde q \tilde \chi^0_1 ) + {\rm BF}( (\tilde q^* \tilde q) \to {\rm SM's} ) \right)  n^2_{\tilde q}
+ \langle \Gamma(\tilde \chi^0_1 \to \tilde q) \rangle 
 \left(n_{\tilde \chi^0_1} - n_{\tilde q}\frac{n^{eq}_{\tilde \chi^0_1}}{n^{eq}_{\tilde q}}\right),\\
\frac{d n_{\tilde \chi^0_1}}{dt} =& -3 H n_{\tilde \chi^0_1} + \frac{ \langle \sigma_{\rm form} v_{\rm rel} \rangle  }{2}  {\rm BF}( (\tilde q^* \tilde q) \to \tilde q \tilde \chi^0_1 )  n^2_{\tilde q} - \langle \Gamma(\tilde \chi^0_1 \to \tilde q) \rangle  \left(n_{\tilde \chi^0_1} - n_{\tilde q}\frac{n^{eq}_{\tilde \chi^0_1}}{n^{eq}_{\tilde q}} \right),
\end{align}
where BF stands for the thermal averaged effective branching fraction of the $\tilde q^* \tilde q$ bound state\,\footnote{Note that we assume that the annihilation between super-hadrons also occurs in the same rate as the annihilation between super-hadron and anti-super-hadron, as discussed in Sec.\,\ref{sec:nonpertann}.}.
We assume the non-perturbative annihilation starts at the QCD phase transition temperature $T=T_c$.
In the following, we take this temperature 200 MeV.
In the QCD confined phase, the degrees of freedom of squark hadrons is non-trivial.
In our estimation, we simply assume that this value is 6 as in the case of a free squark.
In this calculation, we have used values of $g_{*s}$ and $g_*$ estimated in Ref.\,\cite{Saikawa:2018rcs}.
The initial yield is the perturbative value derived in Sec.\,\ref{sec:pertann}. As we have discussed in the previous subsection, the conversion process and thus the end of the integration highly depends on the squark flavor structure and mass difference  and in some case fails to keep the chemical equilibrium.

\begin{figure}[h]
\centering
\includegraphics[clip, width = 0.7 \textwidth]{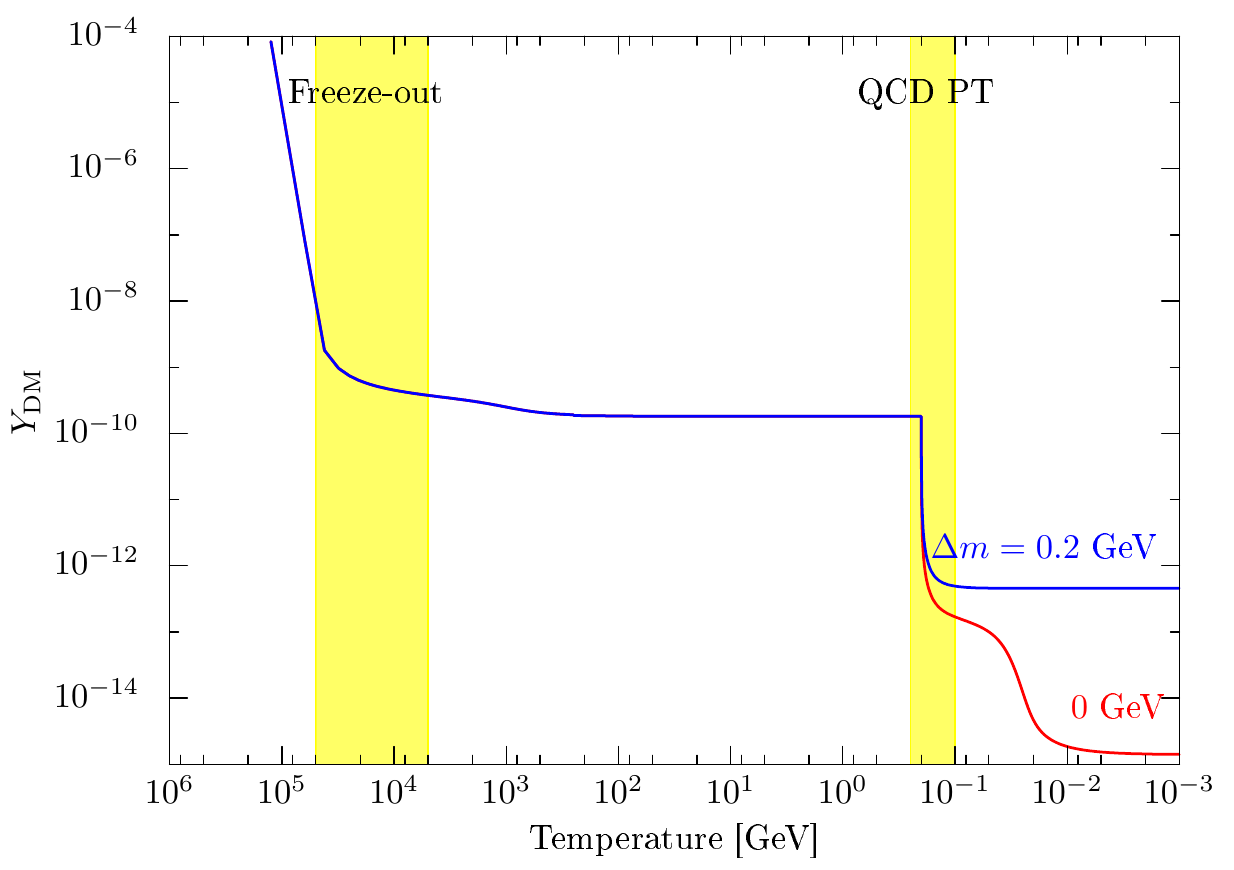} 
\caption{The cosmological evolution of the dark matter yield for $m_{\rm DM} = 10^6$ GeV.
The red (blue) line represent the mass difference with the $\tilde{u}_{R}$ squark $\Delta m$ being 0 (0.2) GeV.
Here we set $\Lambda_{\rm had} = 0.4$ GeV.
}
\label{fig:evolution}
\end{figure}

Finally, we show our result. First, the temperature dependence of the yield for the bino and $\tilde{u}_R$ is shown in Fig.\,\ref{fig:evolution}, where $m_{\rm DM} = 10^6$ GeV and we take $\Lambda_\text{had} = 400$ MeV.
We show the cases of the mass difference $\Delta m = 0$ and 0.2 GeV in red and blue lines, respectively.
We see that,  at $T \sim m_{\rm DM}/30$, the first freeze-out occurs.
When $T=200$ MeV, the non-perturbative annihilation starts and the abundance rapidly reduces.
In the case that $\Delta m = 0$ GeV, at $T\sim 10$ MeV, the annihilation happens further again, since the effective annihilation rate is enhanced as seen in Fig.~\ref{fig:branching_fraction}.

\begin{figure}[h]
\centering
\includegraphics[clip, width = 0.7 \textwidth]{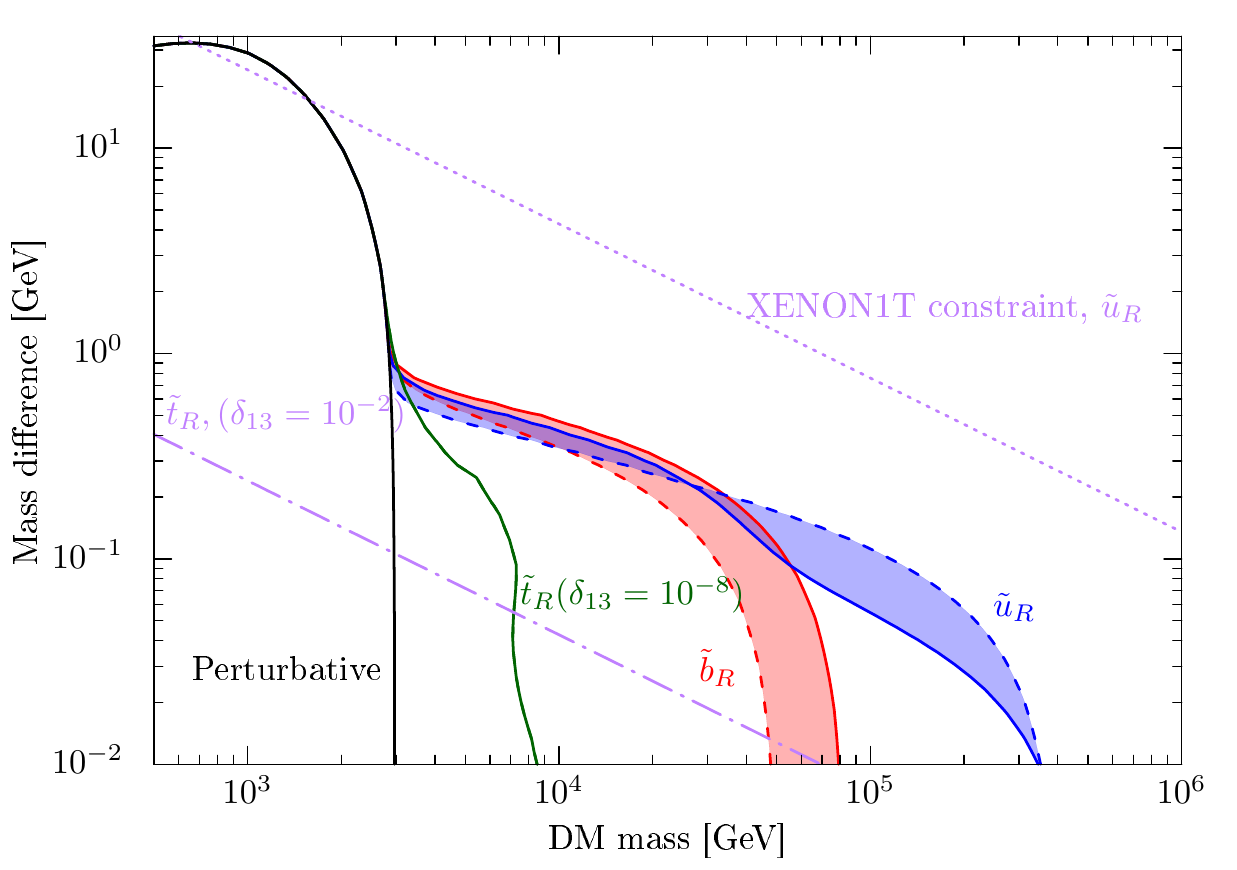} 
\caption{The contour plot of $\Omega_{\rm{DM}} h^2 = 0.12$ on the $m_{\rm DM}$-$\Delta m$ plane.
The black line shows the result of the perturbative estimation from the first freeze-out.
The blue, red and green lines show the cases of $\tilde u_R$, $\tilde b_R$ and $\tilde t_R$, respectively.
For the $\tilde t_R$ we set the mass mixing with the first generation $\delta_{13} = 10^{-8}$.
The solid and dashed lines show the estimation with $\Lambda = 0.4$ and 1 GeV, respectively.
We also plot in dotted purple line the current constraint on the dark matter direct detection with XENON1T \cite{Aprile:2018dbl}, for the case of the bino LSP and $\tilde u_R$ NLSP.
The region below this line is disfavored.
}
\label{fig:m_dm}
\end{figure}

In Fig.~\ref{fig:m_dm}, we show the contour plot of $\Omega_{\rm{DM}} h^2 = 0.12$ \cite{Aghanim:2018eyx} on $m_{\rm DM}-\Delta m$ plane.
We assume the LSP is the bino.
The blue, red and green lines show the cases of $\tilde u_R$, $\tilde b_R$ and $\tilde t_R$ NLSP, respectively.
Without flavor violation, the bino and $\tilde t_R$ are not in chemical equilibrium and we assume a small mixing between the first generation $\delta_{13} = 10^{-8}$, which leads to partial chemical equilibrium.
A small flavor violation $\delta_{13}$ larger than $O(10^{-6})$  can keep almost complete chemical equilibrium.
Moreover the branching fraction of the resonance $(\tilde q^* \tilde q)$ into SM particles also increases (Fig.\,\ref{fig:branching_fraction}) accordingly the dark matter abundance is smaller than the $\tilde u_R$ case.
For the $\tilde b_R$ case, the chemical decoupling occurs at $T\sim 100$ MeV and $\tilde b_R$ cannot annihilate at $T \sim 10$ MeV, where the annihilation cross section is enhanced.
Thus at small mass differences, the abundance for  the $\tilde b_R$ case is larger than the $\tilde u_R$ case.

In the discussions above, we mainly consider the bino LSP.
For the wino and Higgsino LSP, essentially identical estimation is possible.
The DM abundance of these cases is slightly larger than the bino case, due to the large number of degrees of freedom.

Now let us discuss the uncertainties of the present estimation.
As we focus on the non-perturbative QCD phase, there are so many $O(1)$ uncertainties in every stage of the estimation.
One of the largest uncertainties of our calculation is the de-excitation rate of the $\tilde q^* \tilde q$ bound state.
In this analysis, we adopt the estimation based on Refs.\,\cite{Jacoby:2007nw} and \cite{Kusakabe:2011hk}, which indicates $O(10^{-2})$ suppression, compared to the estimation by the naive dimensional analysis.
If there is no such suppression, the DM abundance can be further reduced.

\section{Direct Detection} \label{sec:DD}
\begin{figure}[h]
\centering
\includegraphics[clip, width = 0.7 \textwidth]{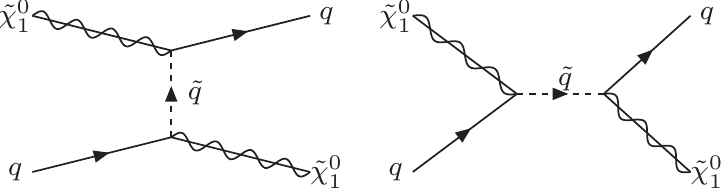} 
\caption{Tree level diagrams for the DM scattering.}
\label{fig:DD}
\end{figure}

In this section, we discuss the direct detection of the dark matter. Since the squark and neutralino LSP are degenerated in mass, the dark matter scattering cross section to the nucleon is enhanced \cite{Drees:1993bu,Hisano:2010ct,Gondolo:2013wwa}. The Feynman diagram is shown in Fig.\,(\ref{fig:DD}). First, we consider bino LSP and right-handed squark NLSP. Then, we briefly discuss the other cases.

The effective Lagrangian for spin-independent scattering between the bino and the quark by the  squark exchange is\,\cite{Hisano:2011um}
\begin{eqnarray}
\mathcal L_\text{eff} = f_q m_q \bar{\tilde{B}}\tilde{B} \bar{q} q + \frac{g_q^{(1)}}{m_{\tilde \chi^0_1}} \bar{\tilde{B}} i \partial^\mu \gamma^\nu \tilde B \mathcal{O}^{q}_{\mu\nu} + \frac{g_q^{(2)}}{m_{\tilde \chi^0_1}^2} \bar{\tilde{B}} i \partial^\mu i \partial^\nu \tilde B \mathcal{O}^{q}_{\mu\nu},
\end{eqnarray}
where $\tilde B$ is the bino fermion field and  $\mathcal{O}^{q}_{\mu\nu} \equiv \frac12 \bar{q} \left(D_\mu \gamma_\nu + D_\nu \gamma_\mu - \frac12 g_{\mu\nu} \slashed{D}\right)q$. The coefficients $f_q$ and $g_q^{(1,2)}$ are given as
\begin{eqnarray}
f_q &\simeq& \frac{c^2 g'^2 m_{\tilde \chi^0_1}}{8} \frac{1}{\left(m_{\tilde{q}}^2 - \left(m_{\tilde \chi^0_1} + m_q\right)^2\right)^2}, \\
g_q^{(1)} &\simeq& \frac{c^2 g'^2 m_{\tilde \chi^0_1}}{2} \frac{1}{\left(m_{\tilde{q}}^2 - \left(m_{\tilde \chi^0_1} + m_q\right)^2\right)^2}
\end{eqnarray}
and $g_q^{(2)} = 0$.
In the present setup, the mixing between the right-handed and left-handed squarks is negligible.

The spin-independent scattering cross section of the bino with a nucleon (atomic number $Z$ and mass number $A$) is given by
\begin{eqnarray}
\sigma_\text{SI} = \frac{4}{\pi} M_\text{red}^2 \left(Z f_p + (A-Z) f_n\right)^2,
\end{eqnarray}
where $M_\text{red}$ is the reduced mass of the bino and nucleon system.
\begin{eqnarray}
\frac{f_N}{m_N} = f_q f_{Tq}^N + \frac34 \left(q(2) + \bar{q}(2)\right) (g_q^{(1)} + g_q^{(2)})
\end{eqnarray}
for a nucleon $N$. The form factor $f_{Tq}^N$ is given by 
\begin{eqnarray}
m_N f_{Tq}^N &\equiv& \langle N|m_q \bar{q}q|N\rangle,
\end{eqnarray}
and $q(2)$ and $\bar{q}(2)$ are the second moments of the parton distribution function.

Using the lattice calculation for $f_{Tq}^N$\,\cite{Abdel-Rehim:2016won} and the CTEQ parton distribution function\,\cite{Kovarik:2015cma} at the $Z$ Boson mass scale, we obtain
\begin{eqnarray}
\sigma_\text{SI} \simeq s_{Ni} \left(\frac{m_{\tilde \chi^0_1}}{10^6~\text{GeV}}\right)^{-2} \left(\frac{\Delta m}{1~\text{GeV} }\right)^{-4} \,\text{cm}^2
\label{eq:DD1}
\end{eqnarray}
for the nucleon $N$ and squark $\tilde{q}_i = \tilde{u}, \tilde{d}$ and $\tilde{s}$. For the heavier flavor $\tilde{q}_i = \tilde{c}$ and $\tilde{b}$, assuming $\Delta m \sim m_{q_i}$,
\begin{eqnarray}
\sigma_\text{SI} \simeq s_{Ni} \left(\frac{m_{\tilde \chi^0_1}}{10^6~\text{GeV}}\right)^{-2} \,\text{cm}^2.
\label{eq:DD2}
\end{eqnarray}
We summarize the coefficient $s_{Ni}$ in Tab.\,\ref{tab:ddc}.
Note that, however, if the mass difference is too small, the estimation at tree-level is not accurate \cite{Drees:1993bu,Hisano:2010ct,Gondolo:2013wwa}.
Moreover in such a case, the uncertainty from the non-perturbative QCD effects will be also sizable.

\begin{table}[htbp]
\centering
\caption{The coefficients for the direct detection in Eqs.\,\eqref{eq:DD1} and \eqref{eq:DD2}. }\label{tab:ddc}
\begin{tabular}{c|c|c|c|c|c}
nucleon & $\tilde{u}$ & $\tilde{d}$ & $\tilde{s}$ & $\tilde{c}$ & $\tilde{b}$ \\\hline
proton & $7.2 \times 10^{-46}$ & $1.6 \times 10^{-47}$ & $3.0 \times 10^{-48}$ & $1.8 \times 10^{-47}$ & $9.5 \times 10^{-52}$ \\
neutron & $2.4 \times 10^{-46}$ & $4.7 \times 10^{-47}$ & $3.0 \times 10^{-48}$ & $1.8 \times 10^{-47}$ & $9.5 \times 10^{-52}$
\end{tabular}
\end{table}

In Fig.~\ref{fig:m_dm}, we show the constraint for the case of $\tilde u_R$ NLSP.
We see that even 100 TeV DM can be excluded in this case.
In fact, the light flavor squark coannihilation cases are also disfavored by the direct detection constraint.
This constraint can be, however, evaded by introducing non-minimal flavor violation.
For instance if the $\tilde t_R$ is the NLSP and there is a very small flavor mixing with the first generation, e.g., $\delta_{13} = 10^{-4}$, we can still keep the successful coannihilation in the QCD era, and avoid the DM direct detection experiments as the cross section is suppressed by a factor of $(\delta_{13})^4$.
A similar estimation also applies to the wino and Higgsino DM.
In the case of the Higgsino DM, the direct detection rate is suppressed by the Yukawa couplings of the quarks and the constraint of the direct detection is also relaxed.

Although the DM detection cross section with very degenerated squark can have large uncertainty, we see that anyway if the mass difference is smaller, the cross section is significantly increased.
The direct detection constraints will disfavor several combinations of the LSP species and squark flavors.

%%%%%%%%%%%%%%%%%%%%%%%%%%%%%%%%%%%%%%%%
\section{Conclusion and Discussion}
\label{sec:conclusion}
%%%%%%%%%%%%%%%%%%%%%%%%%%%%%%%%%%%%%%%%
In this paper, we investigate the possibility that superheavy neutralino dark matter realizes a viable thermal relic density.
If the mass difference between the colored NLSP and the neutralino LSP is smaller than $O(1)$ GeV, non-perturbative QCD effects drastically enhance the NLSP annihilation cross section, and further reduce the DM density.
Although there are large uncertainties in the estimation of the non-perturbative effects, we found that a sub-PeV scale neutralino DM can be consistent with the current observation.
Compared to the previous study for the relic abundance estimation of the colored heavy particles, the destruction of the resonance of two-body bound state of the colored heavy particles is increased and then the effective annihilation cross section of the colored particles is reduced.
Moreover, we study the chemical equilibrium condition between the LSP and NLSP and estimate the effect of out-of-chemical equilibrium on the DM abundance.

Such a highly mass degeneration between the LSP and colored NLSP increases the DM direct detection cross section.
We see that several cases of DM nature and flavor of the squark NLSP are already excluded by the current DM direct detection experiments.
In future direct detection experiments, much larger parameter space, typically ten times larger mass difference than the current constraint can be tested. 

Now let us comment on the gluino coannihilation case.
In principle, the gluino annihilation rate can be increased after the QCD phase transition era.
However, as discussed before,  chemical equilibrium of the gluino and neutralino is prevented if the squark is too heavy.
This constraint implies that the upper-bound of the squark is $O(1-10)$ TeV, depending on the type of the LSP neutralino.
In the gluino coannihilation case, the gluino mass should be less than the squark mass and accordingly the upper-bound of the gluino is also $O(1-10)$ TeV.
We leave the detailed study in a future work \cite{gluinoInprep}.

In this paper, we consider the case that the LSP and NLSP mass is extremely close to each other.
Such a fine-tuned mass spectrum may not be natural.
However, it is very important to clarify what is the maximal DM mass in the SSM, for comprehensive searches of SSMs.

%%%%%%%%%%%%%%%%%%%%%%%%%%%%%%%%%%%%
\section*{Acknowledgments}
%%%%%%%%%%%%%%%%%%%%%%%%%%%%%%%%%%%%
S.S. thanks to A. Kamada and S. Matsumoto for useful discussions.
This work is supported by Grant-in-Aid for Scientific Research from the Ministry of Education, Culture, Sports, Science, and Technology (MEXT), Japan, No. 17H02878 and 18K13535 (S.S.) and  by World Premier International Research Center Initiative (WPI), MEXT, Japan.
The work of H.F. is supported
in part by a Research Fellowship for Young Scientists
from the Japan Society for the Promotion of Science
(JSPS). F.L. is supported by the One Hundred Talent Program of Sun Yat-sen University, China. 

{\it Note Added:} 
During the completion of this paper, a related study
\cite{Gross:2018zha} was submitted to arXiv which discusses the non-perturbative effect of the colored coannihilation.

%%%%%%%%%%%%% References %%%%%%%%%%%%%%%%%%%
\bibliographystyle{aps}
\bibliography{papers}

\begin{thebibliography}{98}%
\makeatletter
\providecommand \@ifxundefined [1]{%
 \@ifx{#1\undefined}
}%
\providecommand \@ifnum [1]{%
 \ifnum #1\expandafter \@firstoftwo
 \else \expandafter \@secondoftwo
 \fi
}%
\providecommand \@ifx [1]{%
 \ifx #1\expandafter \@firstoftwo
 \else \expandafter \@secondoftwo
 \fi
}%
\providecommand \natexlab [1]{#1}%
\providecommand \enquote  [1]{``#1''}%
\providecommand \bibnamefont  [1]{#1}%
\providecommand \bibfnamefont [1]{#1}%
\providecommand \citenamefont [1]{#1}%
\providecommand \href@noop [0]{\@secondoftwo}%
\providecommand \href [0]{\begingroup \@sanitize@url \@href}%
\providecommand \@href[1]{\@@startlink{#1}\@@href}%
\providecommand \@@href[1]{\endgroup#1\@@endlink}%
\providecommand \@sanitize@url [0]{\catcode `\\12\catcode `\$12\catcode
  `\&12\catcode `\#12\catcode `\^12\catcode `\_12\catcode `\%12\relax}%
\providecommand \@@startlink[1]{}%
\providecommand \@@endlink[0]{}%
\providecommand \url  [0]{\begingroup\@sanitize@url \@url }%
\providecommand \@url [1]{\endgroup\@href {#1}{\urlprefix }}%
\providecommand \urlprefix  [0]{URL }%
\providecommand \Eprint [0]{\href }%
\providecommand \doibase [0]{http://dx.doi.org/}%
\providecommand \selectlanguage [0]{\@gobble}%
\providecommand \bibinfo  [0]{\@secondoftwo}%
\providecommand \bibfield  [0]{\@secondoftwo}%
\providecommand \translation [1]{[#1]}%
\providecommand \BibitemOpen [0]{}%
\providecommand \bibitemStop [0]{}%
\providecommand \bibitemNoStop [0]{.\EOS\space}%
\providecommand \EOS [0]{\spacefactor3000\relax}%
\providecommand \BibitemShut  [1]{\csname bibitem#1\endcsname}%
\let\auto@bib@innerbib\@empty
%</preamble>
\bibitem [{\citenamefont {Hisano}\ \emph {et~al.}(2007)\citenamefont {Hisano},
  \citenamefont {Matsumoto}, \citenamefont {Nagai}, \citenamefont {Saito},\
  and\ \citenamefont {Senami}}]{Hisano:2006nn}%
  \BibitemOpen
  \bibfield  {author} {\bibinfo {author} {\bibfnamefont {J.}~\bibnamefont
  {Hisano}}, \bibinfo {author} {\bibfnamefont {S.}~\bibnamefont {Matsumoto}},
  \bibinfo {author} {\bibfnamefont {M.}~\bibnamefont {Nagai}}, \bibinfo
  {author} {\bibfnamefont {O.}~\bibnamefont {Saito}}, \ and\ \bibinfo {author}
  {\bibfnamefont {M.}~\bibnamefont {Senami}},\ }\href {\doibase
  10.1016/j.physletb.2007.01.012} {\bibfield  {journal} {\bibinfo  {journal}
  {Phys. Lett.}\ }\textbf {\bibinfo {volume} {B646}},\ \bibinfo {pages} {34}
  (\bibinfo {year} {2007})},\ \Eprint
  {http://arxiv.org/abs/hep-ph/0610249}{arXiv:hep-ph/0610249
  [hep-ph]}\BibitemShut {NoStop}%
%%CITATION = HEP-PH/0610249;%%
\bibitem [{\citenamefont {Randall}\ and\ \citenamefont
  {Sundrum}(1999)}]{Randall:1998uk}%
  \BibitemOpen
  \bibfield  {author} {\bibinfo {author} {\bibfnamefont {L.}~\bibnamefont
  {Randall}}\ and\ \bibinfo {author} {\bibfnamefont {R.}~\bibnamefont
  {Sundrum}},\ }\href {\doibase 10.1016/S0550-3213(99)00359-4} {\bibfield
  {journal} {\bibinfo  {journal} {Nucl. Phys.}\ }\textbf {\bibinfo {volume}
  {B557}},\ \bibinfo {pages} {79} (\bibinfo {year} {1999})},\ \Eprint
  {http://arxiv.org/abs/hep-th/9810155}{arXiv:hep-th/9810155
  [hep-th]}\BibitemShut {NoStop}%
%%CITATION = HEP-TH/9810155;%%
\bibitem [{\citenamefont {Giudice}\ \emph {et~al.}(1998)\citenamefont
  {Giudice}, \citenamefont {Luty}, \citenamefont {Murayama},\ and\
  \citenamefont {Rattazzi}}]{Giudice:1998xp}%
  \BibitemOpen
  \bibfield  {author} {\bibinfo {author} {\bibfnamefont {G.~F.}\ \bibnamefont
  {Giudice}}, \bibinfo {author} {\bibfnamefont {M.~A.}\ \bibnamefont {Luty}},
  \bibinfo {author} {\bibfnamefont {H.}~\bibnamefont {Murayama}}, \ and\
  \bibinfo {author} {\bibfnamefont {R.}~\bibnamefont {Rattazzi}},\ }\href
  {\doibase 10.1088/1126-6708/1998/12/027} {\bibfield  {journal} {\bibinfo
  {journal} {JHEP}\ }\textbf {\bibinfo {volume} {12}},\ \bibinfo {pages} {027}
  (\bibinfo {year} {1998})},\ \Eprint
  {http://arxiv.org/abs/hep-ph/9810442}{arXiv:hep-ph/9810442
  [hep-ph]}\BibitemShut {NoStop}%
%%CITATION = HEP-PH/9810442;%%
\bibitem [{\citenamefont {Wells}(2003)}]{Wells:2003tf}%
  \BibitemOpen
  \bibfield  {author} {\bibinfo {author} {\bibfnamefont {J.~D.}\ \bibnamefont
  {Wells}},\ }in\ \href@noop {} {\emph {\bibinfo {booktitle} {{11th
  International Conference on Supersymmetry and the Unification of Fundamental
  Interactions (SUSY 2003) Tucson, Arizona, June 5-10, 2003}}}}\ (\bibinfo
  {year} {2003})\ \Eprint
  {http://arxiv.org/abs/hep-ph/0306127}{arXiv:hep-ph/0306127
  [hep-ph]}\BibitemShut {NoStop}%
%%CITATION = HEP-PH/0306127;%%
\bibitem [{\citenamefont {Wells}(2005)}]{Wells:2004di}%
  \BibitemOpen
  \bibfield  {author} {\bibinfo {author} {\bibfnamefont {J.~D.}\ \bibnamefont
  {Wells}},\ }\href {\doibase 10.1103/PhysRevD.71.015013} {\bibfield  {journal}
  {\bibinfo  {journal} {Phys. Rev.}\ }\textbf {\bibinfo {volume} {D71}},\
  \bibinfo {pages} {015013} (\bibinfo {year} {2005})},\ \Eprint
  {http://arxiv.org/abs/hep-ph/0411041}{arXiv:hep-ph/0411041
  [hep-ph]}\BibitemShut {NoStop}%
%%CITATION = HEP-PH/0411041;%%
\bibitem [{\citenamefont {Arkani-Hamed}\ and\ \citenamefont
  {Dimopoulos}(2005)}]{ArkaniHamed:2004fb}%
  \BibitemOpen
  \bibfield  {author} {\bibinfo {author} {\bibfnamefont {N.}~\bibnamefont
  {Arkani-Hamed}}\ and\ \bibinfo {author} {\bibfnamefont {S.}~\bibnamefont
  {Dimopoulos}},\ }\href {\doibase 10.1088/1126-6708/2005/06/073} {\bibfield
  {journal} {\bibinfo  {journal} {JHEP}\ }\textbf {\bibinfo {volume} {06}},\
  \bibinfo {pages} {073} (\bibinfo {year} {2005})},\ \Eprint
  {http://arxiv.org/abs/hep-th/0405159}{arXiv:hep-th/0405159
  [hep-th]}\BibitemShut {NoStop}%
%%CITATION = HEP-TH/0405159;%%
\bibitem [{\citenamefont {Giudice}\ and\ \citenamefont
  {Romanino}(2004)}]{Giudice:2004tc}%
  \BibitemOpen
  \bibfield  {author} {\bibinfo {author} {\bibfnamefont {G.~F.}\ \bibnamefont
  {Giudice}}\ and\ \bibinfo {author} {\bibfnamefont {A.}~\bibnamefont
  {Romanino}},\ }\href {\doibase 10.1016/j.nuclphysb.2004.11.048,
  10.1016/j.nuclphysb.2004.08.001} {\bibfield  {journal} {\bibinfo  {journal}
  {Nucl. Phys.}\ }\textbf {\bibinfo {volume} {B699}},\ \bibinfo {pages} {65}
  (\bibinfo {year} {2004})},\ \bibinfo {note} {[Erratum: Nucl.
  Phys.B706,487(2005)]},\ \Eprint
  {http://arxiv.org/abs/hep-ph/0406088}{arXiv:hep-ph/0406088
  [hep-ph]}\BibitemShut {NoStop}%
%%CITATION = HEP-PH/0406088;%%
\bibitem [{\citenamefont {Arkani-Hamed}\ \emph
  {et~al.}(2005{\natexlab{a}})\citenamefont {Arkani-Hamed}, \citenamefont
  {Dimopoulos}, \citenamefont {Giudice},\ and\ \citenamefont
  {Romanino}}]{ArkaniHamed:2004yi}%
  \BibitemOpen
  \bibfield  {author} {\bibinfo {author} {\bibfnamefont {N.}~\bibnamefont
  {Arkani-Hamed}}, \bibinfo {author} {\bibfnamefont {S.}~\bibnamefont
  {Dimopoulos}}, \bibinfo {author} {\bibfnamefont {G.~F.}\ \bibnamefont
  {Giudice}}, \ and\ \bibinfo {author} {\bibfnamefont {A.}~\bibnamefont
  {Romanino}},\ }\href {\doibase 10.1016/j.nuclphysb.2004.12.026} {\bibfield
  {journal} {\bibinfo  {journal} {Nucl. Phys.}\ }\textbf {\bibinfo {volume}
  {B709}},\ \bibinfo {pages} {3} (\bibinfo {year} {2005}{\natexlab{a}})},\
  \Eprint {http://arxiv.org/abs/hep-ph/0409232}{arXiv:hep-ph/0409232
  [hep-ph]}\BibitemShut {NoStop}%
%%CITATION = HEP-PH/0409232;%%
\bibitem [{\citenamefont {Arkani-Hamed}\ \emph
  {et~al.}(2005{\natexlab{b}})\citenamefont {Arkani-Hamed}, \citenamefont
  {Dimopoulos},\ and\ \citenamefont {Kachru}}]{ArkaniHamed:2005yv}%
  \BibitemOpen
  \bibfield  {author} {\bibinfo {author} {\bibfnamefont {N.}~\bibnamefont
  {Arkani-Hamed}}, \bibinfo {author} {\bibfnamefont {S.}~\bibnamefont
  {Dimopoulos}}, \ and\ \bibinfo {author} {\bibfnamefont {S.}~\bibnamefont
  {Kachru}},\ }\href@noop {} {\  (\bibinfo {year} {2005}{\natexlab{b}})},\
  \Eprint {http://arxiv.org/abs/hep-th/0501082}{arXiv:hep-th/0501082
  [hep-th]}\BibitemShut {NoStop}%
%%CITATION = HEP-TH/0501082;%%
\bibitem [{\citenamefont {Hall}\ and\ \citenamefont
  {Nomura}(2012)}]{Hall:2011jd}%
  \BibitemOpen
  \bibfield  {author} {\bibinfo {author} {\bibfnamefont {L.~J.}\ \bibnamefont
  {Hall}}\ and\ \bibinfo {author} {\bibfnamefont {Y.}~\bibnamefont {Nomura}},\
  }\href {\doibase 10.1007/JHEP01(2012)082} {\bibfield  {journal} {\bibinfo
  {journal} {JHEP}\ }\textbf {\bibinfo {volume} {01}},\ \bibinfo {pages} {082}
  (\bibinfo {year} {2012})},\ \Eprint
  {http://arxiv.org/abs/1111.4519}{arXiv:1111.4519 [hep-ph]}\BibitemShut
  {NoStop}%
%%CITATION = ARXIV:1111.4519;%%
\bibitem [{\citenamefont {Hall}\ \emph {et~al.}(2013)\citenamefont {Hall},
  \citenamefont {Nomura},\ and\ \citenamefont {Shirai}}]{Hall:2012zp}%
  \BibitemOpen
  \bibfield  {author} {\bibinfo {author} {\bibfnamefont {L.~J.}\ \bibnamefont
  {Hall}}, \bibinfo {author} {\bibfnamefont {Y.}~\bibnamefont {Nomura}}, \ and\
  \bibinfo {author} {\bibfnamefont {S.}~\bibnamefont {Shirai}},\ }\href
  {\doibase 10.1007/JHEP01(2013)036} {\bibfield  {journal} {\bibinfo  {journal}
  {JHEP}\ }\textbf {\bibinfo {volume} {01}},\ \bibinfo {pages} {036} (\bibinfo
  {year} {2013})},\ \Eprint {http://arxiv.org/abs/1210.2395}{arXiv:1210.2395
  [hep-ph]}\BibitemShut {NoStop}%
%%CITATION = ARXIV:1210.2395;%%
\bibitem [{\citenamefont {Nomura}\ and\ \citenamefont
  {Shirai}(2014)}]{Nomura:2014asa}%
  \BibitemOpen
  \bibfield  {author} {\bibinfo {author} {\bibfnamefont {Y.}~\bibnamefont
  {Nomura}}\ and\ \bibinfo {author} {\bibfnamefont {S.}~\bibnamefont
  {Shirai}},\ }\href {\doibase 10.1103/PhysRevLett.113.111801} {\bibfield
  {journal} {\bibinfo  {journal} {Phys. Rev. Lett.}\ }\textbf {\bibinfo
  {volume} {113}},\ \bibinfo {pages} {111801} (\bibinfo {year} {2014})},\
  \Eprint {http://arxiv.org/abs/1407.3785}{arXiv:1407.3785
  [hep-ph]}\BibitemShut {NoStop}%
%%CITATION = ARXIV:1407.3785;%%
\bibitem [{\citenamefont {Ibe}\ and\ \citenamefont
  {Yanagida}(2012)}]{Ibe:2011aa}%
  \BibitemOpen
  \bibfield  {author} {\bibinfo {author} {\bibfnamefont {M.}~\bibnamefont
  {Ibe}}\ and\ \bibinfo {author} {\bibfnamefont {T.~T.}\ \bibnamefont
  {Yanagida}},\ }\href {\doibase 10.1016/j.physletb.2012.02.034} {\bibfield
  {journal} {\bibinfo  {journal} {Phys. Lett.}\ }\textbf {\bibinfo {volume}
  {B709}},\ \bibinfo {pages} {374} (\bibinfo {year} {2012})},\ \Eprint
  {http://arxiv.org/abs/1112.2462}{arXiv:1112.2462 [hep-ph]}\BibitemShut
  {NoStop}%
%%CITATION = ARXIV:1112.2462;%%
\bibitem [{\citenamefont {Ibe}\ \emph {et~al.}(2012)\citenamefont {Ibe},
  \citenamefont {Matsumoto},\ and\ \citenamefont {Yanagida}}]{Ibe:2012hu}%
  \BibitemOpen
  \bibfield  {author} {\bibinfo {author} {\bibfnamefont {M.}~\bibnamefont
  {Ibe}}, \bibinfo {author} {\bibfnamefont {S.}~\bibnamefont {Matsumoto}}, \
  and\ \bibinfo {author} {\bibfnamefont {T.~T.}\ \bibnamefont {Yanagida}},\
  }\href {\doibase 10.1103/PhysRevD.85.095011} {\bibfield  {journal} {\bibinfo
  {journal} {Phys. Rev.}\ }\textbf {\bibinfo {volume} {D85}},\ \bibinfo {pages}
  {095011} (\bibinfo {year} {2012})},\ \Eprint
  {http://arxiv.org/abs/1202.2253}{arXiv:1202.2253 [hep-ph]}\BibitemShut
  {NoStop}%
%%CITATION = ARXIV:1202.2253;%%
\bibitem [{\citenamefont {Arvanitaki}\ \emph {et~al.}(2013)\citenamefont
  {Arvanitaki}, \citenamefont {Craig}, \citenamefont {Dimopoulos},\ and\
  \citenamefont {Villadoro}}]{Arvanitaki:2012ps}%
  \BibitemOpen
  \bibfield  {author} {\bibinfo {author} {\bibfnamefont {A.}~\bibnamefont
  {Arvanitaki}}, \bibinfo {author} {\bibfnamefont {N.}~\bibnamefont {Craig}},
  \bibinfo {author} {\bibfnamefont {S.}~\bibnamefont {Dimopoulos}}, \ and\
  \bibinfo {author} {\bibfnamefont {G.}~\bibnamefont {Villadoro}},\ }\href
  {\doibase 10.1007/JHEP02(2013)126} {\bibfield  {journal} {\bibinfo  {journal}
  {JHEP}\ }\textbf {\bibinfo {volume} {02}},\ \bibinfo {pages} {126} (\bibinfo
  {year} {2013})},\ \Eprint {http://arxiv.org/abs/1210.0555}{arXiv:1210.0555
  [hep-ph]}\BibitemShut {NoStop}%
%%CITATION = ARXIV:1210.0555;%%
\bibitem [{\citenamefont {Arkani-Hamed}\ \emph {et~al.}(2012)\citenamefont
  {Arkani-Hamed}, \citenamefont {Gupta}, \citenamefont {Kaplan}, \citenamefont
  {Weiner},\ and\ \citenamefont {Zorawski}}]{ArkaniHamed:2012gw}%
  \BibitemOpen
  \bibfield  {author} {\bibinfo {author} {\bibfnamefont {N.}~\bibnamefont
  {Arkani-Hamed}}, \bibinfo {author} {\bibfnamefont {A.}~\bibnamefont {Gupta}},
  \bibinfo {author} {\bibfnamefont {D.~E.}\ \bibnamefont {Kaplan}}, \bibinfo
  {author} {\bibfnamefont {N.}~\bibnamefont {Weiner}}, \ and\ \bibinfo {author}
  {\bibfnamefont {T.}~\bibnamefont {Zorawski}},\ }\href@noop {} {\  (\bibinfo
  {year} {2012})},\ \Eprint {http://arxiv.org/abs/1212.6971}{arXiv:1212.6971
  [hep-ph]}\BibitemShut {NoStop}%
%%CITATION = ARXIV:1212.6971;%%
\bibitem [{\citenamefont {Hisano}\ \emph {et~al.}(2004)\citenamefont {Hisano},
  \citenamefont {Matsumoto},\ and\ \citenamefont {Nojiri}}]{Hisano:2003ec}%
  \BibitemOpen
  \bibfield  {author} {\bibinfo {author} {\bibfnamefont {J.}~\bibnamefont
  {Hisano}}, \bibinfo {author} {\bibfnamefont {S.}~\bibnamefont {Matsumoto}}, \
  and\ \bibinfo {author} {\bibfnamefont {M.~M.}\ \bibnamefont {Nojiri}},\
  }\href {\doibase 10.1103/PhysRevLett.92.031303} {\bibfield  {journal}
  {\bibinfo  {journal} {Phys. Rev. Lett.}\ }\textbf {\bibinfo {volume} {92}},\
  \bibinfo {pages} {031303} (\bibinfo {year} {2004})},\ \Eprint
  {http://arxiv.org/abs/hep-ph/0307216}{arXiv:hep-ph/0307216
  [hep-ph]}\BibitemShut {NoStop}%
%%CITATION = HEP-PH/0307216;%%
\bibitem [{\citenamefont {Hisano}\ \emph {et~al.}(2005)\citenamefont {Hisano},
  \citenamefont {Matsumoto}, \citenamefont {Nojiri},\ and\ \citenamefont
  {Saito}}]{Hisano:2004ds}%
  \BibitemOpen
  \bibfield  {author} {\bibinfo {author} {\bibfnamefont {J.}~\bibnamefont
  {Hisano}}, \bibinfo {author} {\bibfnamefont {S.}~\bibnamefont {Matsumoto}},
  \bibinfo {author} {\bibfnamefont {M.~M.}\ \bibnamefont {Nojiri}}, \ and\
  \bibinfo {author} {\bibfnamefont {O.}~\bibnamefont {Saito}},\ }\href
  {\doibase 10.1103/PhysRevD.71.063528} {\bibfield  {journal} {\bibinfo
  {journal} {Phys. Rev.}\ }\textbf {\bibinfo {volume} {D71}},\ \bibinfo {pages}
  {063528} (\bibinfo {year} {2005})},\ \Eprint
  {http://arxiv.org/abs/hep-ph/0412403}{arXiv:hep-ph/0412403
  [hep-ph]}\BibitemShut {NoStop}%
%%CITATION = HEP-PH/0412403;%%
\bibitem [{\citenamefont {Hisano}\ \emph {et~al.}(2006)\citenamefont {Hisano},
  \citenamefont {Matsumoto}, \citenamefont {Saito},\ and\ \citenamefont
  {Senami}}]{Hisano:2005ec}%
  \BibitemOpen
  \bibfield  {author} {\bibinfo {author} {\bibfnamefont {J.}~\bibnamefont
  {Hisano}}, \bibinfo {author} {\bibfnamefont {S.}~\bibnamefont {Matsumoto}},
  \bibinfo {author} {\bibfnamefont {O.}~\bibnamefont {Saito}}, \ and\ \bibinfo
  {author} {\bibfnamefont {M.}~\bibnamefont {Senami}},\ }\href {\doibase
  10.1103/PhysRevD.73.055004} {\bibfield  {journal} {\bibinfo  {journal} {Phys.
  Rev.}\ }\textbf {\bibinfo {volume} {D73}},\ \bibinfo {pages} {055004}
  (\bibinfo {year} {2006})},\ \Eprint
  {http://arxiv.org/abs/hep-ph/0511118}{arXiv:hep-ph/0511118
  [hep-ph]}\BibitemShut {NoStop}%
%%CITATION = HEP-PH/0511118;%%
\bibitem [{\citenamefont {Hisano}\ \emph
  {et~al.}(2010{\natexlab{a}})\citenamefont {Hisano}, \citenamefont
  {Ishiwata},\ and\ \citenamefont {Nagata}}]{Hisano:2010fy}%
  \BibitemOpen
  \bibfield  {author} {\bibinfo {author} {\bibfnamefont {J.}~\bibnamefont
  {Hisano}}, \bibinfo {author} {\bibfnamefont {K.}~\bibnamefont {Ishiwata}}, \
  and\ \bibinfo {author} {\bibfnamefont {N.}~\bibnamefont {Nagata}},\ }\href
  {\doibase 10.1016/j.physletb.2010.05.047} {\bibfield  {journal} {\bibinfo
  {journal} {Phys. Lett.}\ }\textbf {\bibinfo {volume} {B690}},\ \bibinfo
  {pages} {311} (\bibinfo {year} {2010}{\natexlab{a}})},\ \Eprint
  {http://arxiv.org/abs/1004.4090}{arXiv:1004.4090 [hep-ph]}\BibitemShut
  {NoStop}%
%%CITATION = ARXIV:1004.4090;%%
\bibitem [{\citenamefont {Hisano}\ \emph
  {et~al.}(2010{\natexlab{b}})\citenamefont {Hisano}, \citenamefont
  {Ishiwata},\ and\ \citenamefont {Nagata}}]{Hisano:2010ct}%
  \BibitemOpen
  \bibfield  {author} {\bibinfo {author} {\bibfnamefont {J.}~\bibnamefont
  {Hisano}}, \bibinfo {author} {\bibfnamefont {K.}~\bibnamefont {Ishiwata}}, \
  and\ \bibinfo {author} {\bibfnamefont {N.}~\bibnamefont {Nagata}},\ }\href
  {\doibase 10.1103/PhysRevD.82.115007} {\bibfield  {journal} {\bibinfo
  {journal} {Phys. Rev.}\ }\textbf {\bibinfo {volume} {D82}},\ \bibinfo {pages}
  {115007} (\bibinfo {year} {2010}{\natexlab{b}})},\ \Eprint
  {http://arxiv.org/abs/1007.2601}{arXiv:1007.2601 [hep-ph]}\BibitemShut
  {NoStop}%
%%CITATION = ARXIV:1007.2601;%%
\bibitem [{\citenamefont {Hisano}\ \emph {et~al.}(2013)\citenamefont {Hisano},
  \citenamefont {Ishiwata},\ and\ \citenamefont {Nagata}}]{Hisano:2012wm}%
  \BibitemOpen
  \bibfield  {author} {\bibinfo {author} {\bibfnamefont {J.}~\bibnamefont
  {Hisano}}, \bibinfo {author} {\bibfnamefont {K.}~\bibnamefont {Ishiwata}}, \
  and\ \bibinfo {author} {\bibfnamefont {N.}~\bibnamefont {Nagata}},\ }\href
  {\doibase 10.1103/PhysRevD.87.035020} {\bibfield  {journal} {\bibinfo
  {journal} {Phys. Rev.}\ }\textbf {\bibinfo {volume} {D87}},\ \bibinfo {pages}
  {035020} (\bibinfo {year} {2013})},\ \Eprint
  {http://arxiv.org/abs/1210.5985}{arXiv:1210.5985 [hep-ph]}\BibitemShut
  {NoStop}%
%%CITATION = ARXIV:1210.5985;%%
\bibitem [{\citenamefont {Hisano}\ \emph {et~al.}(2015)\citenamefont {Hisano},
  \citenamefont {Ishiwata},\ and\ \citenamefont {Nagata}}]{Hisano:2015rsa}%
  \BibitemOpen
  \bibfield  {author} {\bibinfo {author} {\bibfnamefont {J.}~\bibnamefont
  {Hisano}}, \bibinfo {author} {\bibfnamefont {K.}~\bibnamefont {Ishiwata}}, \
  and\ \bibinfo {author} {\bibfnamefont {N.}~\bibnamefont {Nagata}},\ }\href
  {\doibase 10.1007/JHEP06(2015)097} {\bibfield  {journal} {\bibinfo  {journal}
  {JHEP}\ }\textbf {\bibinfo {volume} {06}},\ \bibinfo {pages} {097} (\bibinfo
  {year} {2015})},\ \Eprint {http://arxiv.org/abs/1504.00915}{arXiv:1504.00915
  [hep-ph]}\BibitemShut {NoStop}%
%%CITATION = ARXIV:1504.00915;%%
\bibitem [{\citenamefont {Ibe}\ \emph {et~al.}(2007)\citenamefont {Ibe},
  \citenamefont {Moroi},\ and\ \citenamefont {Yanagida}}]{Ibe:2006de}%
  \BibitemOpen
  \bibfield  {author} {\bibinfo {author} {\bibfnamefont {M.}~\bibnamefont
  {Ibe}}, \bibinfo {author} {\bibfnamefont {T.}~\bibnamefont {Moroi}}, \ and\
  \bibinfo {author} {\bibfnamefont {T.~T.}\ \bibnamefont {Yanagida}},\ }\href
  {\doibase 10.1016/j.physletb.2006.11.061} {\bibfield  {journal} {\bibinfo
  {journal} {Phys. Lett.}\ }\textbf {\bibinfo {volume} {B644}},\ \bibinfo
  {pages} {355} (\bibinfo {year} {2007})},\ \Eprint
  {http://arxiv.org/abs/hep-ph/0610277}{arXiv:hep-ph/0610277
  [hep-ph]}\BibitemShut {NoStop}%
%%CITATION = HEP-PH/0610277;%%
\bibitem [{\citenamefont {Buckley}\ \emph {et~al.}(2011)\citenamefont
  {Buckley}, \citenamefont {Randall},\ and\ \citenamefont
  {Shuve}}]{Buckley:2009kv}%
  \BibitemOpen
  \bibfield  {author} {\bibinfo {author} {\bibfnamefont {M.~R.}\ \bibnamefont
  {Buckley}}, \bibinfo {author} {\bibfnamefont {L.}~\bibnamefont {Randall}}, \
  and\ \bibinfo {author} {\bibfnamefont {B.}~\bibnamefont {Shuve}},\ }\href
  {\doibase 10.1007/JHEP05(2011)097} {\bibfield  {journal} {\bibinfo  {journal}
  {JHEP}\ }\textbf {\bibinfo {volume} {05}},\ \bibinfo {pages} {097} (\bibinfo
  {year} {2011})},\ \Eprint {http://arxiv.org/abs/0909.4549}{arXiv:0909.4549
  [hep-ph]}\BibitemShut {NoStop}%
%%CITATION = ARXIV:0909.4549;%%
\bibitem [{\citenamefont {Asai}\ \emph {et~al.}(2007)\citenamefont {Asai},
  \citenamefont {Moroi}, \citenamefont {Nishihara},\ and\ \citenamefont
  {Yanagida}}]{Asai:2007sw}%
  \BibitemOpen
  \bibfield  {author} {\bibinfo {author} {\bibfnamefont {S.}~\bibnamefont
  {Asai}}, \bibinfo {author} {\bibfnamefont {T.}~\bibnamefont {Moroi}},
  \bibinfo {author} {\bibfnamefont {K.}~\bibnamefont {Nishihara}}, \ and\
  \bibinfo {author} {\bibfnamefont {T.~T.}\ \bibnamefont {Yanagida}},\ }\href
  {\doibase 10.1016/j.physletb.2007.06.080} {\bibfield  {journal} {\bibinfo
  {journal} {Phys. Lett.}\ }\textbf {\bibinfo {volume} {B653}},\ \bibinfo
  {pages} {81} (\bibinfo {year} {2007})},\ \Eprint
  {http://arxiv.org/abs/0705.3086}{arXiv:0705.3086 [hep-ph]}\BibitemShut
  {NoStop}%
%%CITATION = ARXIV:0705.3086;%%
\bibitem [{\citenamefont {Asai}\ \emph {et~al.}(2008)\citenamefont {Asai},
  \citenamefont {Moroi},\ and\ \citenamefont {Yanagida}}]{Asai:2008sk}%
  \BibitemOpen
  \bibfield  {author} {\bibinfo {author} {\bibfnamefont {S.}~\bibnamefont
  {Asai}}, \bibinfo {author} {\bibfnamefont {T.}~\bibnamefont {Moroi}}, \ and\
  \bibinfo {author} {\bibfnamefont {T.~T.}\ \bibnamefont {Yanagida}},\ }\href
  {\doibase 10.1016/j.physletb.2008.05.019} {\bibfield  {journal} {\bibinfo
  {journal} {Phys. Lett.}\ }\textbf {\bibinfo {volume} {B664}},\ \bibinfo
  {pages} {185} (\bibinfo {year} {2008})},\ \Eprint
  {http://arxiv.org/abs/0802.3725}{arXiv:0802.3725 [hep-ph]}\BibitemShut
  {NoStop}%
%%CITATION = ARXIV:0802.3725;%%
\bibitem [{\citenamefont {Asai}\ \emph {et~al.}(2009)\citenamefont {Asai},
  \citenamefont {Azuma}, \citenamefont {Jinnouchi}, \citenamefont {Moroi},
  \citenamefont {Shirai},\ and\ \citenamefont {Yanagida}}]{Asai:2008im}%
  \BibitemOpen
  \bibfield  {author} {\bibinfo {author} {\bibfnamefont {S.}~\bibnamefont
  {Asai}}, \bibinfo {author} {\bibfnamefont {Y.}~\bibnamefont {Azuma}},
  \bibinfo {author} {\bibfnamefont {O.}~\bibnamefont {Jinnouchi}}, \bibinfo
  {author} {\bibfnamefont {T.}~\bibnamefont {Moroi}}, \bibinfo {author}
  {\bibfnamefont {S.}~\bibnamefont {Shirai}}, \ and\ \bibinfo {author}
  {\bibfnamefont {T.~T.}\ \bibnamefont {Yanagida}},\ }\href {\doibase
  10.1016/j.physletb.2009.01.045} {\bibfield  {journal} {\bibinfo  {journal}
  {Phys. Lett.}\ }\textbf {\bibinfo {volume} {B672}},\ \bibinfo {pages} {339}
  (\bibinfo {year} {2009})},\ \Eprint
  {http://arxiv.org/abs/0807.4987}{arXiv:0807.4987 [hep-ph]}\BibitemShut
  {NoStop}%
%%CITATION = ARXIV:0807.4987;%%
\bibitem [{\citenamefont {Mahbubani}\ \emph {et~al.}(2017)\citenamefont
  {Mahbubani}, \citenamefont {Schwaller},\ and\ \citenamefont
  {Zurita}}]{Mahbubani:2017gjh}%
  \BibitemOpen
  \bibfield  {author} {\bibinfo {author} {\bibfnamefont {R.}~\bibnamefont
  {Mahbubani}}, \bibinfo {author} {\bibfnamefont {P.}~\bibnamefont
  {Schwaller}}, \ and\ \bibinfo {author} {\bibfnamefont {J.}~\bibnamefont
  {Zurita}},\ }\href {\doibase 10.1007/JHEP06(2017)119,
  10.1007/JHEP10(2017)061} {\bibfield  {journal} {\bibinfo  {journal} {JHEP}\
  }\textbf {\bibinfo {volume} {06}},\ \bibinfo {pages} {119} (\bibinfo {year}
  {2017})},\ \bibinfo {note} {[Erratum: JHEP10,061(2017)]},\ \Eprint
  {http://arxiv.org/abs/1703.05327}{arXiv:1703.05327 [hep-ph]}\BibitemShut
  {NoStop}%
%%CITATION = ARXIV:1703.05327;%%
\bibitem [{\citenamefont {Nagata}\ and\ \citenamefont
  {Shirai}(2015)}]{Nagata:2014wma}%
  \BibitemOpen
  \bibfield  {author} {\bibinfo {author} {\bibfnamefont {N.}~\bibnamefont
  {Nagata}}\ and\ \bibinfo {author} {\bibfnamefont {S.}~\bibnamefont
  {Shirai}},\ }\href {\doibase 10.1007/JHEP01(2015)029} {\bibfield  {journal}
  {\bibinfo  {journal} {JHEP}\ }\textbf {\bibinfo {volume} {01}},\ \bibinfo
  {pages} {029} (\bibinfo {year} {2015})},\ \Eprint
  {http://arxiv.org/abs/1410.4549}{arXiv:1410.4549 [hep-ph]}\BibitemShut
  {NoStop}%
%%CITATION = ARXIV:1410.4549;%%
\bibitem [{\citenamefont {Fukuda}\ \emph {et~al.}(2017)\citenamefont {Fukuda},
  \citenamefont {Nagata}, \citenamefont {Otono},\ and\ \citenamefont
  {Shirai}}]{Fukuda:2017jmk}%
  \BibitemOpen
  \bibfield  {author} {\bibinfo {author} {\bibfnamefont {H.}~\bibnamefont
  {Fukuda}}, \bibinfo {author} {\bibfnamefont {N.}~\bibnamefont {Nagata}},
  \bibinfo {author} {\bibfnamefont {H.}~\bibnamefont {Otono}}, \ and\ \bibinfo
  {author} {\bibfnamefont {S.}~\bibnamefont {Shirai}},\ }\href@noop {} {\
  (\bibinfo {year} {2017})},\ \Eprint
  {http://arxiv.org/abs/1703.09675}{arXiv:1703.09675 [hep-ph]}\BibitemShut
  {NoStop}%
%%CITATION = ARXIV:1703.09675;%%
\bibitem [{\citenamefont {Matsumoto}\ \emph
  {et~al.}(2018{\natexlab{a}})\citenamefont {Matsumoto}, \citenamefont
  {Shirai},\ and\ \citenamefont {Takeuchi}}]{Matsumoto:2017vfu}%
  \BibitemOpen
  \bibfield  {author} {\bibinfo {author} {\bibfnamefont {S.}~\bibnamefont
  {Matsumoto}}, \bibinfo {author} {\bibfnamefont {S.}~\bibnamefont {Shirai}}, \
  and\ \bibinfo {author} {\bibfnamefont {M.}~\bibnamefont {Takeuchi}},\ }\href
  {\doibase 10.1007/JHEP06(2018)049} {\bibfield  {journal} {\bibinfo  {journal}
  {JHEP}\ }\textbf {\bibinfo {volume} {06}},\ \bibinfo {pages} {049} (\bibinfo
  {year} {2018}{\natexlab{a}})},\ \Eprint
  {http://arxiv.org/abs/1711.05449}{arXiv:1711.05449 [hep-ph]}\BibitemShut
  {NoStop}%
%%CITATION = ARXIV:1711.05449;%%
\bibitem [{\citenamefont {Matsumoto}\ \emph
  {et~al.}(2018{\natexlab{b}})\citenamefont {Matsumoto}, \citenamefont
  {Shirai},\ and\ \citenamefont {Takeuchi}}]{Matsumoto:2018ioi}%
  \BibitemOpen
  \bibfield  {author} {\bibinfo {author} {\bibfnamefont {S.}~\bibnamefont
  {Matsumoto}}, \bibinfo {author} {\bibfnamefont {S.}~\bibnamefont {Shirai}}, \
  and\ \bibinfo {author} {\bibfnamefont {M.}~\bibnamefont {Takeuchi}},\
  }\href@noop {} {\  (\bibinfo {year} {2018}{\natexlab{b}})},\ \Eprint
  {http://arxiv.org/abs/1810.12234}{arXiv:1810.12234 [hep-ph]}\BibitemShut
  {NoStop}%
%%CITATION = ARXIV:1810.12234;%%
\bibitem [{\citenamefont {Chigusa}\ \emph {et~al.}(2018)\citenamefont
  {Chigusa}, \citenamefont {Ema},\ and\ \citenamefont
  {Moroi}}]{Chigusa:2018vxz}%
  \BibitemOpen
  \bibfield  {author} {\bibinfo {author} {\bibfnamefont {S.}~\bibnamefont
  {Chigusa}}, \bibinfo {author} {\bibfnamefont {Y.}~\bibnamefont {Ema}}, \ and\
  \bibinfo {author} {\bibfnamefont {T.}~\bibnamefont {Moroi}},\ }\href@noop {}
  {\  (\bibinfo {year} {2018})},\ \Eprint
  {http://arxiv.org/abs/1810.07349}{arXiv:1810.07349 [hep-ph]}\BibitemShut
  {NoStop}%
%%CITATION = ARXIV:1810.07349;%%
\bibitem [{\citenamefont {Di~Luzio}\ \emph {et~al.}(2018)\citenamefont
  {Di~Luzio}, \citenamefont {Gröber},\ and\ \citenamefont
  {Panico}}]{DiLuzio:2018jwd}%
  \BibitemOpen
  \bibfield  {author} {\bibinfo {author} {\bibfnamefont {L.}~\bibnamefont
  {Di~Luzio}}, \bibinfo {author} {\bibfnamefont {R.}~\bibnamefont {Gröber}}, \
  and\ \bibinfo {author} {\bibfnamefont {G.}~\bibnamefont {Panico}},\
  }\href@noop {} {\  (\bibinfo {year} {2018})},\ \Eprint
  {http://arxiv.org/abs/1810.10993}{arXiv:1810.10993 [hep-ph]}\BibitemShut
  {NoStop}%
%%CITATION = ARXIV:1810.10993;%%
\bibitem [{\citenamefont {Griest}\ and\ \citenamefont
  {Seckel}(1991)}]{Griest:1990kh}%
  \BibitemOpen
  \bibfield  {author} {\bibinfo {author} {\bibfnamefont {K.}~\bibnamefont
  {Griest}}\ and\ \bibinfo {author} {\bibfnamefont {D.}~\bibnamefont
  {Seckel}},\ }\href {\doibase 10.1103/PhysRevD.43.3191} {\bibfield  {journal}
  {\bibinfo  {journal} {Phys. Rev.}\ }\textbf {\bibinfo {volume} {D43}},\
  \bibinfo {pages} {3191} (\bibinfo {year} {1991})}\BibitemShut {NoStop}%
%%CITATION = PHRVA,D43,3191;%%
\bibitem [{\citenamefont {Edsjo}\ and\ \citenamefont
  {Gondolo}(1997)}]{Edsjo:1997bg}%
  \BibitemOpen
  \bibfield  {author} {\bibinfo {author} {\bibfnamefont {J.}~\bibnamefont
  {Edsjo}}\ and\ \bibinfo {author} {\bibfnamefont {P.}~\bibnamefont
  {Gondolo}},\ }\href {\doibase 10.1103/PhysRevD.56.1879} {\bibfield  {journal}
  {\bibinfo  {journal} {Phys. Rev.}\ }\textbf {\bibinfo {volume} {D56}},\
  \bibinfo {pages} {1879} (\bibinfo {year} {1997})},\ \Eprint
  {http://arxiv.org/abs/hep-ph/9704361}{arXiv:hep-ph/9704361
  [hep-ph]}\BibitemShut {NoStop}%
%%CITATION = HEP-PH/9704361;%%
\bibitem [{\citenamefont {Profumo}(2005)}]{Profumo:2005xd}%
  \BibitemOpen
  \bibfield  {author} {\bibinfo {author} {\bibfnamefont {S.}~\bibnamefont
  {Profumo}},\ }\href {\doibase 10.1103/PhysRevD.72.103521} {\bibfield
  {journal} {\bibinfo  {journal} {Phys. Rev.}\ }\textbf {\bibinfo {volume}
  {D72}},\ \bibinfo {pages} {103521} (\bibinfo {year} {2005})},\ \Eprint
  {http://arxiv.org/abs/astro-ph/0508628}{arXiv:astro-ph/0508628
  [astro-ph]}\BibitemShut {NoStop}%
%%CITATION = ASTRO-PH/0508628;%%
\bibitem [{\citenamefont {Gilmore}(2007)}]{Gilmore:2007aq}%
  \BibitemOpen
  \bibfield  {author} {\bibinfo {author} {\bibfnamefont {R.~C.}\ \bibnamefont
  {Gilmore}},\ }\href {\doibase 10.1103/PhysRevD.76.043520} {\bibfield
  {journal} {\bibinfo  {journal} {Phys. Rev.}\ }\textbf {\bibinfo {volume}
  {D76}},\ \bibinfo {pages} {043520} (\bibinfo {year} {2007})},\ \Eprint
  {http://arxiv.org/abs/0705.2610}{arXiv:0705.2610 [hep-ph]}\BibitemShut
  {NoStop}%
%%CITATION = ARXIV:0705.2610;%%
\bibitem [{\citenamefont {Fukuda}\ and\ \citenamefont {Shirai}()}]{resInprep}%
  \BibitemOpen
  \bibfield  {author} {\bibinfo {author} {\bibfnamefont {H.}~\bibnamefont
  {Fukuda}}\ and\ \bibinfo {author} {\bibfnamefont {S.}~\bibnamefont
  {Shirai}},\ }\href@noop {} {\bibinfo  {journal} {to appear}\ }\BibitemShut
  {NoStop}%
\bibitem [{\citenamefont {Harigaya}\ \emph {et~al.}(2014)\citenamefont
  {Harigaya}, \citenamefont {Kaneta},\ and\ \citenamefont
  {Matsumoto}}]{Harigaya:2014dwa}%
  \BibitemOpen
\bibfield  {journal} {  }\bibfield  {author} {\bibinfo {author} {\bibfnamefont
  {K.}~\bibnamefont {Harigaya}}, \bibinfo {author} {\bibfnamefont
  {K.}~\bibnamefont {Kaneta}}, \ and\ \bibinfo {author} {\bibfnamefont
  {S.}~\bibnamefont {Matsumoto}},\ }\href {\doibase 10.1103/PhysRevD.89.115021}
  {\bibfield  {journal} {\bibinfo  {journal} {Phys. Rev.}\ }\textbf {\bibinfo
  {volume} {D89}},\ \bibinfo {pages} {115021} (\bibinfo {year} {2014})},\
  \Eprint {http://arxiv.org/abs/1403.0715}{arXiv:1403.0715
  [hep-ph]}\BibitemShut {NoStop}%
%%CITATION = ARXIV:1403.0715;%%
\bibitem [{\citenamefont {Ellis}\ \emph {et~al.}(2015)\citenamefont {Ellis},
  \citenamefont {Luo},\ and\ \citenamefont {Olive}}]{Ellis:2015vaa}%
  \BibitemOpen
  \bibfield  {author} {\bibinfo {author} {\bibfnamefont {J.}~\bibnamefont
  {Ellis}}, \bibinfo {author} {\bibfnamefont {F.}~\bibnamefont {Luo}}, \ and\
  \bibinfo {author} {\bibfnamefont {K.~A.}\ \bibnamefont {Olive}},\ }\href
  {\doibase 10.1007/JHEP09(2015)127} {\bibfield  {journal} {\bibinfo  {journal}
  {JHEP}\ }\textbf {\bibinfo {volume} {09}},\ \bibinfo {pages} {127} (\bibinfo
  {year} {2015})},\ \Eprint {http://arxiv.org/abs/1503.07142}{arXiv:1503.07142
  [hep-ph]}\BibitemShut {NoStop}%
%%CITATION = ARXIV:1503.07142;%%
\bibitem [{\citenamefont {Ellis}\ \emph {et~al.}(2016)\citenamefont {Ellis},
  \citenamefont {Evans}, \citenamefont {Luo},\ and\ \citenamefont
  {Olive}}]{Ellis:2015vna}%
  \BibitemOpen
  \bibfield  {author} {\bibinfo {author} {\bibfnamefont {J.}~\bibnamefont
  {Ellis}}, \bibinfo {author} {\bibfnamefont {J.~L.}\ \bibnamefont {Evans}},
  \bibinfo {author} {\bibfnamefont {F.}~\bibnamefont {Luo}}, \ and\ \bibinfo
  {author} {\bibfnamefont {K.~A.}\ \bibnamefont {Olive}},\ }\href {\doibase
  10.1007/JHEP02(2016)071} {\bibfield  {journal} {\bibinfo  {journal} {JHEP}\
  }\textbf {\bibinfo {volume} {02}},\ \bibinfo {pages} {071} (\bibinfo {year}
  {2016})},\ \Eprint {http://arxiv.org/abs/1510.03498}{arXiv:1510.03498
  [hep-ph]}\BibitemShut {NoStop}%
%%CITATION = ARXIV:1510.03498;%%
\bibitem [{\citenamefont {Liew}\ and\ \citenamefont
  {Luo}(2017)}]{Liew:2016hqo}%
  \BibitemOpen
  \bibfield  {author} {\bibinfo {author} {\bibfnamefont {S.~P.}\ \bibnamefont
  {Liew}}\ and\ \bibinfo {author} {\bibfnamefont {F.}~\bibnamefont {Luo}},\
  }\href {\doibase 10.1007/JHEP02(2017)091} {\bibfield  {journal} {\bibinfo
  {journal} {JHEP}\ }\textbf {\bibinfo {volume} {02}},\ \bibinfo {pages} {091}
  (\bibinfo {year} {2017})},\ \Eprint
  {http://arxiv.org/abs/1611.08133}{arXiv:1611.08133 [hep-ph]}\BibitemShut
  {NoStop}%
%%CITATION = ARXIV:1611.08133;%%
\bibitem [{\citenamefont {Nagata}\ \emph {et~al.}(2015)\citenamefont {Nagata},
  \citenamefont {Otono},\ and\ \citenamefont {Shirai}}]{Nagata:2015hha}%
  \BibitemOpen
  \bibfield  {author} {\bibinfo {author} {\bibfnamefont {N.}~\bibnamefont
  {Nagata}}, \bibinfo {author} {\bibfnamefont {H.}~\bibnamefont {Otono}}, \
  and\ \bibinfo {author} {\bibfnamefont {S.}~\bibnamefont {Shirai}},\ }\href
  {\doibase 10.1016/j.physletb.2015.06.044} {\bibfield  {journal} {\bibinfo
  {journal} {Phys. Lett.}\ }\textbf {\bibinfo {volume} {B748}},\ \bibinfo
  {pages} {24} (\bibinfo {year} {2015})},\ \Eprint
  {http://arxiv.org/abs/1504.00504}{arXiv:1504.00504 [hep-ph]}\BibitemShut
  {NoStop}%
%%CITATION = ARXIV:1504.00504;%%
\bibitem [{\citenamefont {Nagata}\ \emph {et~al.}(2017)\citenamefont {Nagata},
  \citenamefont {Otono},\ and\ \citenamefont {Shirai}}]{Nagata:2017gci}%
  \BibitemOpen
  \bibfield  {author} {\bibinfo {author} {\bibfnamefont {N.}~\bibnamefont
  {Nagata}}, \bibinfo {author} {\bibfnamefont {H.}~\bibnamefont {Otono}}, \
  and\ \bibinfo {author} {\bibfnamefont {S.}~\bibnamefont {Shirai}},\ }\href
  {\doibase 10.1007/JHEP03(2017)025} {\bibfield  {journal} {\bibinfo  {journal}
  {JHEP}\ }\textbf {\bibinfo {volume} {03}},\ \bibinfo {pages} {025} (\bibinfo
  {year} {2017})},\ \Eprint {http://arxiv.org/abs/1701.07664}{arXiv:1701.07664
  [hep-ph]}\BibitemShut {NoStop}%
%%CITATION = ARXIV:1701.07664;%%
\bibitem [{\citenamefont {Ellis}\ and\ \citenamefont
  {Zheng}(2015)}]{Ellis:2015xba}%
  \BibitemOpen
  \bibfield  {author} {\bibinfo {author} {\bibfnamefont {S.~A.~R.}\
  \bibnamefont {Ellis}}\ and\ \bibinfo {author} {\bibfnamefont
  {B.}~\bibnamefont {Zheng}},\ }\href {\doibase 10.1103/PhysRevD.92.075034}
  {\bibfield  {journal} {\bibinfo  {journal} {Phys. Rev.}\ }\textbf {\bibinfo
  {volume} {D92}},\ \bibinfo {pages} {075034} (\bibinfo {year} {2015})},\
  \Eprint {http://arxiv.org/abs/1506.02644}{arXiv:1506.02644
  [hep-ph]}\BibitemShut {NoStop}%
%%CITATION = ARXIV:1506.02644;%%
\bibitem [{\citenamefont {Griest}\ and\ \citenamefont
  {Kamionkowski}(1990)}]{Griest:1989wd}%
  \BibitemOpen
  \bibfield  {author} {\bibinfo {author} {\bibfnamefont {K.}~\bibnamefont
  {Griest}}\ and\ \bibinfo {author} {\bibfnamefont {M.}~\bibnamefont
  {Kamionkowski}},\ }\href {\doibase 10.1103/PhysRevLett.64.615} {\bibfield
  {journal} {\bibinfo  {journal} {Phys. Rev. Lett.}\ }\textbf {\bibinfo
  {volume} {64}},\ \bibinfo {pages} {615} (\bibinfo {year} {1990})}\BibitemShut
  {NoStop}%
%%CITATION = PRLTA,64,615;%%
\bibitem [{\citenamefont {Harigaya}\ \emph {et~al.}(2016)\citenamefont
  {Harigaya}, \citenamefont {Ibe}, \citenamefont {Kaneta}, \citenamefont
  {Nakano},\ and\ \citenamefont {Suzuki}}]{Harigaya:2016nlg}%
  \BibitemOpen
  \bibfield  {author} {\bibinfo {author} {\bibfnamefont {K.}~\bibnamefont
  {Harigaya}}, \bibinfo {author} {\bibfnamefont {M.}~\bibnamefont {Ibe}},
  \bibinfo {author} {\bibfnamefont {K.}~\bibnamefont {Kaneta}}, \bibinfo
  {author} {\bibfnamefont {W.}~\bibnamefont {Nakano}}, \ and\ \bibinfo {author}
  {\bibfnamefont {M.}~\bibnamefont {Suzuki}},\ }\href {\doibase
  10.1007/JHEP08(2016)151} {\bibfield  {journal} {\bibinfo  {journal} {JHEP}\
  }\textbf {\bibinfo {volume} {08}},\ \bibinfo {pages} {151} (\bibinfo {year}
  {2016})},\ \Eprint {http://arxiv.org/abs/1606.00159}{arXiv:1606.00159
  [hep-ph]}\BibitemShut {NoStop}%
%%CITATION = ARXIV:1606.00159;%%
\bibitem [{\citenamefont {Ellis}\ \emph {et~al.}(2018)\citenamefont {Ellis},
  \citenamefont {Evans}, \citenamefont {Luo}, \citenamefont {Olive},\ and\
  \citenamefont {Zheng}}]{Ellis:2018jyl}%
  \BibitemOpen
  \bibfield  {author} {\bibinfo {author} {\bibfnamefont {J.}~\bibnamefont
  {Ellis}}, \bibinfo {author} {\bibfnamefont {J.~L.}\ \bibnamefont {Evans}},
  \bibinfo {author} {\bibfnamefont {F.}~\bibnamefont {Luo}}, \bibinfo {author}
  {\bibfnamefont {K.~A.}\ \bibnamefont {Olive}}, \ and\ \bibinfo {author}
  {\bibfnamefont {J.}~\bibnamefont {Zheng}},\ }\href {\doibase
  10.1140/epjc/s10052-018-5831-z} {\bibfield  {journal} {\bibinfo  {journal}
  {Eur. Phys. J.}\ }\textbf {\bibinfo {volume} {C78}},\ \bibinfo {pages} {425}
  (\bibinfo {year} {2018})},\ \Eprint
  {http://arxiv.org/abs/1801.09855}{arXiv:1801.09855 [hep-ph]}\BibitemShut
  {NoStop}%
%%CITATION = ARXIV:1801.09855;%%
\bibitem [{\citenamefont {De~Simone}\ \emph {et~al.}(2014)\citenamefont
  {De~Simone}, \citenamefont {Giudice},\ and\ \citenamefont
  {Strumia}}]{deSimone:2014pda}%
  \BibitemOpen
  \bibfield  {author} {\bibinfo {author} {\bibfnamefont {A.}~\bibnamefont
  {De~Simone}}, \bibinfo {author} {\bibfnamefont {G.~F.}\ \bibnamefont
  {Giudice}}, \ and\ \bibinfo {author} {\bibfnamefont {A.}~\bibnamefont
  {Strumia}},\ }\href {\doibase 10.1007/JHEP06(2014)081} {\bibfield  {journal}
  {\bibinfo  {journal} {JHEP}\ }\textbf {\bibinfo {volume} {06}},\ \bibinfo
  {pages} {081} (\bibinfo {year} {2014})},\ \Eprint
  {http://arxiv.org/abs/1402.6287}{arXiv:1402.6287 [hep-ph]}\BibitemShut
  {NoStop}%
%%CITATION = ARXIV:1402.6287;%%
\bibitem [{\citenamefont {Ellis}\ \emph {et~al.}(2014)\citenamefont {Ellis},
  \citenamefont {Olive},\ and\ \citenamefont {Zheng}}]{Ellis:2014ipa}%
  \BibitemOpen
  \bibfield  {author} {\bibinfo {author} {\bibfnamefont {J.}~\bibnamefont
  {Ellis}}, \bibinfo {author} {\bibfnamefont {K.~A.}\ \bibnamefont {Olive}}, \
  and\ \bibinfo {author} {\bibfnamefont {J.}~\bibnamefont {Zheng}},\ }\href
  {\doibase 10.1140/epjc/s10052-014-2947-7} {\bibfield  {journal} {\bibinfo
  {journal} {Eur. Phys. J.}\ }\textbf {\bibinfo {volume} {C74}},\ \bibinfo
  {pages} {2947} (\bibinfo {year} {2014})},\ \Eprint
  {http://arxiv.org/abs/1404.5571}{arXiv:1404.5571 [hep-ph]}\BibitemShut
  {NoStop}%
%%CITATION = ARXIV:1404.5571;%%
\bibitem [{\citenamefont {von Harling}\ and\ \citenamefont
  {Petraki}(2014)}]{vonHarling:2014kha}%
  \BibitemOpen
  \bibfield  {author} {\bibinfo {author} {\bibfnamefont {B.}~\bibnamefont {von
  Harling}}\ and\ \bibinfo {author} {\bibfnamefont {K.}~\bibnamefont
  {Petraki}},\ }\href {\doibase 10.1088/1475-7516/2014/12/033} {\bibfield
  {journal} {\bibinfo  {journal} {JCAP}\ }\textbf {\bibinfo {volume} {1412}},\
  \bibinfo {pages} {033} (\bibinfo {year} {2014})},\ \Eprint
  {http://arxiv.org/abs/1407.7874}{arXiv:1407.7874 [hep-ph]}\BibitemShut
  {NoStop}%
%%CITATION = ARXIV:1407.7874;%%
\bibitem [{\citenamefont {Ibarra}\ \emph {et~al.}(2015)\citenamefont {Ibarra},
  \citenamefont {Pierce}, \citenamefont {Shah},\ and\ \citenamefont
  {Vogl}}]{Ibarra:2015nca}%
  \BibitemOpen
  \bibfield  {author} {\bibinfo {author} {\bibfnamefont {A.}~\bibnamefont
  {Ibarra}}, \bibinfo {author} {\bibfnamefont {A.}~\bibnamefont {Pierce}},
  \bibinfo {author} {\bibfnamefont {N.~R.}\ \bibnamefont {Shah}}, \ and\
  \bibinfo {author} {\bibfnamefont {S.}~\bibnamefont {Vogl}},\ }\bibfield
  {booktitle} {\emph {\bibinfo {booktitle} {{Proceedings, Meeting of the APS
  Division of Particles and Fields (DPF 2015): Ann Arbor, Michigan, USA, 4-8
  Aug 2015}}},\ }\href {\doibase 10.1103/PhysRevD.91.095018} {\bibfield
  {journal} {\bibinfo  {journal} {Phys. Rev.}\ }\textbf {\bibinfo {volume}
  {D91}},\ \bibinfo {pages} {095018} (\bibinfo {year} {2015})},\ \Eprint
  {http://arxiv.org/abs/1501.03164}{arXiv:1501.03164 [hep-ph]}\BibitemShut
  {NoStop}%
%%CITATION = ARXIV:1501.03164;%%
\bibitem [{\citenamefont {Cirelli}\ \emph {et~al.}(2017)\citenamefont
  {Cirelli}, \citenamefont {Panci}, \citenamefont {Petraki}, \citenamefont
  {Sala},\ and\ \citenamefont {Taoso}}]{Cirelli:2016rnw}%
  \BibitemOpen
  \bibfield  {author} {\bibinfo {author} {\bibfnamefont {M.}~\bibnamefont
  {Cirelli}}, \bibinfo {author} {\bibfnamefont {P.}~\bibnamefont {Panci}},
  \bibinfo {author} {\bibfnamefont {K.}~\bibnamefont {Petraki}}, \bibinfo
  {author} {\bibfnamefont {F.}~\bibnamefont {Sala}}, \ and\ \bibinfo {author}
  {\bibfnamefont {M.}~\bibnamefont {Taoso}},\ }\href {\doibase
  10.1088/1475-7516/2017/05/036} {\bibfield  {journal} {\bibinfo  {journal}
  {JCAP}\ }\textbf {\bibinfo {volume} {1705}},\ \bibinfo {pages} {036}
  (\bibinfo {year} {2017})},\ \Eprint
  {http://arxiv.org/abs/1612.07295}{arXiv:1612.07295 [hep-ph]}\BibitemShut
  {NoStop}%
%%CITATION = ARXIV:1612.07295;%%
\bibitem [{\citenamefont {Kim}\ and\ \citenamefont
  {Laine}(2016)}]{Kim:2016zyy}%
  \BibitemOpen
  \bibfield  {author} {\bibinfo {author} {\bibfnamefont {S.}~\bibnamefont
  {Kim}}\ and\ \bibinfo {author} {\bibfnamefont {M.}~\bibnamefont {Laine}},\
  }\href {\doibase 10.1007/JHEP07(2016)143} {\bibfield  {journal} {\bibinfo
  {journal} {JHEP}\ }\textbf {\bibinfo {volume} {07}},\ \bibinfo {pages} {143}
  (\bibinfo {year} {2016})},\ \Eprint
  {http://arxiv.org/abs/1602.08105}{arXiv:1602.08105 [hep-ph]}\BibitemShut
  {NoStop}%
%%CITATION = ARXIV:1602.08105;%%
\bibitem [{\citenamefont {Kim}\ and\ \citenamefont
  {Laine}(2017)}]{Kim:2016kxt}%
  \BibitemOpen
  \bibfield  {author} {\bibinfo {author} {\bibfnamefont {S.}~\bibnamefont
  {Kim}}\ and\ \bibinfo {author} {\bibfnamefont {M.}~\bibnamefont {Laine}},\
  }\href {\doibase 10.1088/1475-7516/2017/01/013} {\bibfield  {journal}
  {\bibinfo  {journal} {JCAP}\ }\textbf {\bibinfo {volume} {1701}},\ \bibinfo
  {pages} {013} (\bibinfo {year} {2017})},\ \Eprint
  {http://arxiv.org/abs/1609.00474}{arXiv:1609.00474 [hep-ph]}\BibitemShut
  {NoStop}%
%%CITATION = ARXIV:1609.00474;%%
\bibitem [{\citenamefont {El~Hedri}\ \emph
  {et~al.}(2017{\natexlab{a}})\citenamefont {El~Hedri}, \citenamefont
  {Kaminska},\ and\ \citenamefont {de~Vries}}]{ElHedri:2016onc}%
  \BibitemOpen
  \bibfield  {author} {\bibinfo {author} {\bibfnamefont {S.}~\bibnamefont
  {El~Hedri}}, \bibinfo {author} {\bibfnamefont {A.}~\bibnamefont {Kaminska}},
  \ and\ \bibinfo {author} {\bibfnamefont {M.}~\bibnamefont {de~Vries}},\
  }\href {\doibase 10.1140/epjc/s10052-017-5168-z} {\bibfield  {journal}
  {\bibinfo  {journal} {Eur. Phys. J.}\ }\textbf {\bibinfo {volume} {C77}},\
  \bibinfo {pages} {622} (\bibinfo {year} {2017}{\natexlab{a}})},\ \Eprint
  {http://arxiv.org/abs/1612.02825}{arXiv:1612.02825 [hep-ph]}\BibitemShut
  {NoStop}%
%%CITATION = ARXIV:1612.02825;%%
\bibitem [{\citenamefont {Petraki}\ \emph {et~al.}(2017)\citenamefont
  {Petraki}, \citenamefont {Postma},\ and\ \citenamefont
  {de~Vries}}]{Petraki:2016cnz}%
  \BibitemOpen
  \bibfield  {author} {\bibinfo {author} {\bibfnamefont {K.}~\bibnamefont
  {Petraki}}, \bibinfo {author} {\bibfnamefont {M.}~\bibnamefont {Postma}}, \
  and\ \bibinfo {author} {\bibfnamefont {J.}~\bibnamefont {de~Vries}},\ }\href
  {\doibase 10.1007/JHEP04(2017)077} {\bibfield  {journal} {\bibinfo  {journal}
  {JHEP}\ }\textbf {\bibinfo {volume} {04}},\ \bibinfo {pages} {077} (\bibinfo
  {year} {2017})},\ \Eprint {http://arxiv.org/abs/1611.01394}{arXiv:1611.01394
  [hep-ph]}\BibitemShut {NoStop}%
%%CITATION = ARXIV:1611.01394;%%
\bibitem [{\citenamefont {El~Hedri}\ \emph
  {et~al.}(2017{\natexlab{b}})\citenamefont {El~Hedri}, \citenamefont
  {Kaminska}, \citenamefont {de~Vries},\ and\ \citenamefont
  {Zurita}}]{ElHedri:2017nny}%
  \BibitemOpen
  \bibfield  {author} {\bibinfo {author} {\bibfnamefont {S.}~\bibnamefont
  {El~Hedri}}, \bibinfo {author} {\bibfnamefont {A.}~\bibnamefont {Kaminska}},
  \bibinfo {author} {\bibfnamefont {M.}~\bibnamefont {de~Vries}}, \ and\
  \bibinfo {author} {\bibfnamefont {J.}~\bibnamefont {Zurita}},\ }\href
  {\doibase 10.1007/JHEP04(2017)118} {\bibfield  {journal} {\bibinfo  {journal}
  {JHEP}\ }\textbf {\bibinfo {volume} {04}},\ \bibinfo {pages} {118} (\bibinfo
  {year} {2017}{\natexlab{b}})},\ \Eprint
  {http://arxiv.org/abs/1703.00452}{arXiv:1703.00452 [hep-ph]}\BibitemShut
  {NoStop}%
%%CITATION = ARXIV:1703.00452;%%
\bibitem [{\citenamefont {Pierce}\ \emph {et~al.}(2018)\citenamefont {Pierce},
  \citenamefont {Shah},\ and\ \citenamefont {Vogl}}]{Pierce:2017suq}%
  \BibitemOpen
  \bibfield  {author} {\bibinfo {author} {\bibfnamefont {A.}~\bibnamefont
  {Pierce}}, \bibinfo {author} {\bibfnamefont {N.~R.}\ \bibnamefont {Shah}}, \
  and\ \bibinfo {author} {\bibfnamefont {S.}~\bibnamefont {Vogl}},\ }\href
  {\doibase 10.1103/PhysRevD.97.023008} {\bibfield  {journal} {\bibinfo
  {journal} {Phys. Rev.}\ }\textbf {\bibinfo {volume} {D97}},\ \bibinfo {pages}
  {023008} (\bibinfo {year} {2018})},\ \Eprint
  {http://arxiv.org/abs/1706.01911}{arXiv:1706.01911 [hep-ph]}\BibitemShut
  {NoStop}%
%%CITATION = ARXIV:1706.01911;%%
\bibitem [{\citenamefont {Keung}\ \emph {et~al.}(2017)\citenamefont {Keung},
  \citenamefont {Low},\ and\ \citenamefont {Zhang}}]{Keung:2017kot}%
  \BibitemOpen
  \bibfield  {author} {\bibinfo {author} {\bibfnamefont {W.-Y.}\ \bibnamefont
  {Keung}}, \bibinfo {author} {\bibfnamefont {I.}~\bibnamefont {Low}}, \ and\
  \bibinfo {author} {\bibfnamefont {Y.}~\bibnamefont {Zhang}},\ }\href
  {\doibase 10.1103/PhysRevD.96.015008} {\bibfield  {journal} {\bibinfo
  {journal} {Phys. Rev.}\ }\textbf {\bibinfo {volume} {D96}},\ \bibinfo {pages}
  {015008} (\bibinfo {year} {2017})},\ \Eprint
  {http://arxiv.org/abs/1703.02977}{arXiv:1703.02977 [hep-ph]}\BibitemShut
  {NoStop}%
%%CITATION = ARXIV:1703.02977;%%
\bibitem [{\citenamefont {Garny}\ \emph {et~al.}(2017)\citenamefont {Garny},
  \citenamefont {Heisig}, \citenamefont {Lülf},\ and\ \citenamefont
  {Vogl}}]{Garny:2017rxs}%
  \BibitemOpen
  \bibfield  {author} {\bibinfo {author} {\bibfnamefont {M.}~\bibnamefont
  {Garny}}, \bibinfo {author} {\bibfnamefont {J.}~\bibnamefont {Heisig}},
  \bibinfo {author} {\bibfnamefont {B.}~\bibnamefont {Lülf}}, \ and\ \bibinfo
  {author} {\bibfnamefont {S.}~\bibnamefont {Vogl}},\ }\href {\doibase
  10.1103/PhysRevD.96.103521} {\bibfield  {journal} {\bibinfo  {journal} {Phys.
  Rev.}\ }\textbf {\bibinfo {volume} {D96}},\ \bibinfo {pages} {103521}
  (\bibinfo {year} {2017})},\ \Eprint
  {http://arxiv.org/abs/1705.09292}{arXiv:1705.09292 [hep-ph]}\BibitemShut
  {NoStop}%
%%CITATION = ARXIV:1705.09292;%%
\bibitem [{\citenamefont {Harz}\ and\ \citenamefont
  {Petraki}(2018{\natexlab{a}})}]{Harz:2017dlj}%
  \BibitemOpen
  \bibfield  {author} {\bibinfo {author} {\bibfnamefont {J.}~\bibnamefont
  {Harz}}\ and\ \bibinfo {author} {\bibfnamefont {K.}~\bibnamefont {Petraki}},\
  }\href {\doibase 10.1103/PhysRevD.97.075041} {\bibfield  {journal} {\bibinfo
  {journal} {Phys. Rev.}\ }\textbf {\bibinfo {volume} {D97}},\ \bibinfo {pages}
  {075041} (\bibinfo {year} {2018}{\natexlab{a}})},\ \Eprint
  {http://arxiv.org/abs/1711.03552}{arXiv:1711.03552 [hep-ph]}\BibitemShut
  {NoStop}%
%%CITATION = ARXIV:1711.03552;%%
\bibitem [{\citenamefont {Biondini}\ and\ \citenamefont
  {Laine}(2018)}]{Biondini:2018pwp}%
  \BibitemOpen
  \bibfield  {author} {\bibinfo {author} {\bibfnamefont {S.}~\bibnamefont
  {Biondini}}\ and\ \bibinfo {author} {\bibfnamefont {M.}~\bibnamefont
  {Laine}},\ }\href {\doibase 10.1007/JHEP04(2018)072} {\bibfield  {journal}
  {\bibinfo  {journal} {JHEP}\ }\textbf {\bibinfo {volume} {04}},\ \bibinfo
  {pages} {072} (\bibinfo {year} {2018})},\ \Eprint
  {http://arxiv.org/abs/1801.05821}{arXiv:1801.05821 [hep-ph]}\BibitemShut
  {NoStop}%
%%CITATION = ARXIV:1801.05821;%%
\bibitem [{\citenamefont {El~Hedri}\ and\ \citenamefont
  {de~Vries}(2018)}]{ElHedri:2018atj}%
  \BibitemOpen
  \bibfield  {author} {\bibinfo {author} {\bibfnamefont {S.}~\bibnamefont
  {El~Hedri}}\ and\ \bibinfo {author} {\bibfnamefont {M.}~\bibnamefont
  {de~Vries}},\ }\href {\doibase 10.1007/JHEP10(2018)102} {\bibfield  {journal}
  {\bibinfo  {journal} {JHEP}\ }\textbf {\bibinfo {volume} {10}},\ \bibinfo
  {pages} {102} (\bibinfo {year} {2018})},\ \Eprint
  {http://arxiv.org/abs/1806.03325}{arXiv:1806.03325 [hep-ph]}\BibitemShut
  {NoStop}%
%%CITATION = ARXIV:1806.03325;%%
\bibitem [{\citenamefont {Binder}\ \emph {et~al.}(2018)\citenamefont {Binder},
  \citenamefont {Covi},\ and\ \citenamefont {Mukaida}}]{Binder:2018znk}%
  \BibitemOpen
  \bibfield  {author} {\bibinfo {author} {\bibfnamefont {T.}~\bibnamefont
  {Binder}}, \bibinfo {author} {\bibfnamefont {L.}~\bibnamefont {Covi}}, \ and\
  \bibinfo {author} {\bibfnamefont {K.}~\bibnamefont {Mukaida}},\ }\href@noop
  {} {\  (\bibinfo {year} {2018})},\ \Eprint
  {http://arxiv.org/abs/1808.06472}{arXiv:1808.06472 [hep-ph]}\BibitemShut
  {NoStop}%
%%CITATION = ARXIV:1808.06472;%%
\bibitem [{\citenamefont {Biondini}(2018)}]{Biondini:2018xor}%
  \BibitemOpen
  \bibfield  {author} {\bibinfo {author} {\bibfnamefont {S.}~\bibnamefont
  {Biondini}},\ }\href {\doibase 10.1007/JHEP06(2018)104} {\bibfield  {journal}
  {\bibinfo  {journal} {JHEP}\ }\textbf {\bibinfo {volume} {06}},\ \bibinfo
  {pages} {104} (\bibinfo {year} {2018})},\ \Eprint
  {http://arxiv.org/abs/1805.00353}{arXiv:1805.00353 [hep-ph]}\BibitemShut
  {NoStop}%
%%CITATION = ARXIV:1805.00353;%%
\bibitem [{\citenamefont {Biondini}\ and\ \citenamefont
  {Vogl}(2018)}]{Biondini:2018ovz}%
  \BibitemOpen
  \bibfield  {author} {\bibinfo {author} {\bibfnamefont {S.}~\bibnamefont
  {Biondini}}\ and\ \bibinfo {author} {\bibfnamefont {S.}~\bibnamefont
  {Vogl}},\ }\href@noop {} {\  (\bibinfo {year} {2018})},\ \Eprint
  {http://arxiv.org/abs/1811.02581}{arXiv:1811.02581 [hep-ph]}\BibitemShut
  {NoStop}%
%%CITATION = ARXIV:1811.02581;%%
\bibitem [{\citenamefont {Harz}\ and\ \citenamefont
  {Petraki}(2019)}]{Harz:2019rro}%
  \BibitemOpen
  \bibfield  {author} {\bibinfo {author} {\bibfnamefont {J.}~\bibnamefont
  {Harz}}\ and\ \bibinfo {author} {\bibfnamefont {K.}~\bibnamefont {Petraki}},\
  }\href@noop {} {\  (\bibinfo {year} {2019})},\ \Eprint
  {http://arxiv.org/abs/1901.10030}{arXiv:1901.10030 [hep-ph]}\BibitemShut
  {NoStop}%
%%CITATION = ARXIV:1901.10030;%%
\bibitem [{\citenamefont {Mitridate}\ \emph {et~al.}(2017)\citenamefont
  {Mitridate}, \citenamefont {Redi}, \citenamefont {Smirnov},\ and\
  \citenamefont {Strumia}}]{Mitridate:2017izz}%
  \BibitemOpen
  \bibfield  {author} {\bibinfo {author} {\bibfnamefont {A.}~\bibnamefont
  {Mitridate}}, \bibinfo {author} {\bibfnamefont {M.}~\bibnamefont {Redi}},
  \bibinfo {author} {\bibfnamefont {J.}~\bibnamefont {Smirnov}}, \ and\
  \bibinfo {author} {\bibfnamefont {A.}~\bibnamefont {Strumia}},\ }\href
  {\doibase 10.1088/1475-7516/2017/05/006} {\bibfield  {journal} {\bibinfo
  {journal} {JCAP}\ }\textbf {\bibinfo {volume} {1705}},\ \bibinfo {pages}
  {006} (\bibinfo {year} {2017})},\ \Eprint
  {http://arxiv.org/abs/1702.01141}{arXiv:1702.01141 [hep-ph]}\BibitemShut
  {NoStop}%
%%CITATION = ARXIV:1702.01141;%%
\bibitem [{\citenamefont {Harz}\ and\ \citenamefont
  {Petraki}(2018{\natexlab{b}})}]{Harz:2018csl}%
  \BibitemOpen
  \bibfield  {author} {\bibinfo {author} {\bibfnamefont {J.}~\bibnamefont
  {Harz}}\ and\ \bibinfo {author} {\bibfnamefont {K.}~\bibnamefont {Petraki}},\
  }\href {\doibase 10.1007/JHEP07(2018)096} {\bibfield  {journal} {\bibinfo
  {journal} {JHEP}\ }\textbf {\bibinfo {volume} {07}},\ \bibinfo {pages} {096}
  (\bibinfo {year} {2018}{\natexlab{b}})},\ \Eprint
  {http://arxiv.org/abs/1805.01200}{arXiv:1805.01200 [hep-ph]}\BibitemShut
  {NoStop}%
%%CITATION = ARXIV:1805.01200;%%
\bibitem [{\citenamefont {Srednicki}\ \emph {et~al.}(1988)\citenamefont
  {Srednicki}, \citenamefont {Watkins},\ and\ \citenamefont
  {Olive}}]{Srednicki:1988ce}%
  \BibitemOpen
  \bibfield  {author} {\bibinfo {author} {\bibfnamefont {M.}~\bibnamefont
  {Srednicki}}, \bibinfo {author} {\bibfnamefont {R.}~\bibnamefont {Watkins}},
  \ and\ \bibinfo {author} {\bibfnamefont {K.~A.}\ \bibnamefont {Olive}},\
  }\href {\doibase 10.1016/0550-3213(88)90099-5} {\bibfield  {journal}
  {\bibinfo  {journal} {Nucl. Phys.}\ }\textbf {\bibinfo {volume} {B310}},\
  \bibinfo {pages} {693} (\bibinfo {year} {1988})}\BibitemShut {NoStop}%
%%CITATION = NUPHA,B310,693;%%
\bibitem [{\citenamefont {Asadi}\ \emph {et~al.}(2017)\citenamefont {Asadi},
  \citenamefont {Baumgart}, \citenamefont {Fitzpatrick}, \citenamefont
  {Krupczak},\ and\ \citenamefont {Slatyer}}]{Asadi:2016ybp}%
  \BibitemOpen
  \bibfield  {author} {\bibinfo {author} {\bibfnamefont {P.}~\bibnamefont
  {Asadi}}, \bibinfo {author} {\bibfnamefont {M.}~\bibnamefont {Baumgart}},
  \bibinfo {author} {\bibfnamefont {P.~J.}\ \bibnamefont {Fitzpatrick}},
  \bibinfo {author} {\bibfnamefont {E.}~\bibnamefont {Krupczak}}, \ and\
  \bibinfo {author} {\bibfnamefont {T.~R.}\ \bibnamefont {Slatyer}},\ }\href
  {\doibase 10.1088/1475-7516/2017/02/005} {\bibfield  {journal} {\bibinfo
  {journal} {JCAP}\ }\textbf {\bibinfo {volume} {1702}},\ \bibinfo {pages}
  {005} (\bibinfo {year} {2017})},\ \Eprint
  {http://arxiv.org/abs/1610.07617}{arXiv:1610.07617 [hep-ph]}\BibitemShut
  {NoStop}%
%%CITATION = ARXIV:1610.07617;%%
\bibitem [{\citenamefont {Kang}\ \emph {et~al.}(2008)\citenamefont {Kang},
  \citenamefont {Luty},\ and\ \citenamefont {Nasri}}]{Kang:2006yd}%
  \BibitemOpen
  \bibfield  {author} {\bibinfo {author} {\bibfnamefont {J.}~\bibnamefont
  {Kang}}, \bibinfo {author} {\bibfnamefont {M.~A.}\ \bibnamefont {Luty}}, \
  and\ \bibinfo {author} {\bibfnamefont {S.}~\bibnamefont {Nasri}},\ }\href
  {\doibase 10.1088/1126-6708/2008/09/086} {\bibfield  {journal} {\bibinfo
  {journal} {JHEP}\ }\textbf {\bibinfo {volume} {09}},\ \bibinfo {pages} {086}
  (\bibinfo {year} {2008})},\ \Eprint
  {http://arxiv.org/abs/hep-ph/0611322}{arXiv:hep-ph/0611322
  [hep-ph]}\BibitemShut {NoStop}%
%%CITATION = HEP-PH/0611322;%%
\bibitem [{\citenamefont {Geller}\ \emph {et~al.}(2018)\citenamefont {Geller},
  \citenamefont {Iwamoto}, \citenamefont {Lee}, \citenamefont {Shadmi},\ and\
  \citenamefont {Telem}}]{Geller:2018biy}%
  \BibitemOpen
  \bibfield  {author} {\bibinfo {author} {\bibfnamefont {M.}~\bibnamefont
  {Geller}}, \bibinfo {author} {\bibfnamefont {S.}~\bibnamefont {Iwamoto}},
  \bibinfo {author} {\bibfnamefont {G.}~\bibnamefont {Lee}}, \bibinfo {author}
  {\bibfnamefont {Y.}~\bibnamefont {Shadmi}}, \ and\ \bibinfo {author}
  {\bibfnamefont {O.}~\bibnamefont {Telem}},\ }\href {\doibase
  10.1007/JHEP06(2018)135} {\bibfield  {journal} {\bibinfo  {journal} {JHEP}\
  }\textbf {\bibinfo {volume} {06}},\ \bibinfo {pages} {135} (\bibinfo {year}
  {2018})},\ \Eprint {http://arxiv.org/abs/1802.07720}{arXiv:1802.07720
  [hep-ph]}\BibitemShut {NoStop}%
%%CITATION = ARXIV:1802.07720;%%
\bibitem [{\citenamefont {Jacoby}\ and\ \citenamefont
  {Nussinov}(2007)}]{Jacoby:2007nw}%
  \BibitemOpen
  \bibfield  {author} {\bibinfo {author} {\bibfnamefont {C.}~\bibnamefont
  {Jacoby}}\ and\ \bibinfo {author} {\bibfnamefont {S.}~\bibnamefont
  {Nussinov}},\ }\href@noop {} {\  (\bibinfo {year} {2007})},\ \Eprint
  {http://arxiv.org/abs/0712.2681}{arXiv:0712.2681 [hep-ph]}\BibitemShut
  {NoStop}%
%%CITATION = ARXIV:0712.2681;%%
\bibitem [{\citenamefont {Sakurai}(1993)}]{SakuraiQM}%
  \BibitemOpen
  \bibfield  {author} {\bibinfo {author} {\bibfnamefont {J.~J.}\ \bibnamefont
  {Sakurai}},\ }\href@noop {} {\emph {\bibinfo {title} {Modern Quantum
  Mechanics, Revised Edition}}},\ \bibinfo {edition} {1st}\ ed.\ (\bibinfo
  {publisher} {Addison Wesley},\ \bibinfo {year} {1993})\BibitemShut {NoStop}%
\bibitem [{\citenamefont {Tanabashi}\ \emph {et~al.}(2018)\citenamefont
  {Tanabashi} \emph {et~al.}}]{Tanabashi:2018oca}%
  \BibitemOpen
  \bibfield  {author} {\bibinfo {author} {\bibfnamefont {M.}~\bibnamefont
  {Tanabashi}} \emph {et~al.} (\bibinfo {collaboration} {Particle Data
  Group}),\ }\href {\doibase 10.1103/PhysRevD.98.030001} {\bibfield  {journal}
  {\bibinfo  {journal} {Phys. Rev.}\ }\textbf {\bibinfo {volume} {D98}},\
  \bibinfo {pages} {030001} (\bibinfo {year} {2018})}\BibitemShut {NoStop}%
%%CITATION = PHRVA,D98,030001;%%
\bibitem [{\citenamefont {Kusakabe}\ and\ \citenamefont
  {Takesako}(2012)}]{Kusakabe:2011hk}%
  \BibitemOpen
  \bibfield  {author} {\bibinfo {author} {\bibfnamefont {M.}~\bibnamefont
  {Kusakabe}}\ and\ \bibinfo {author} {\bibfnamefont {T.}~\bibnamefont
  {Takesako}},\ }\href {\doibase 10.1103/PhysRevD.85.015005} {\bibfield
  {journal} {\bibinfo  {journal} {Phys. Rev.}\ }\textbf {\bibinfo {volume}
  {D85}},\ \bibinfo {pages} {015005} (\bibinfo {year} {2012})},\ \Eprint
  {http://arxiv.org/abs/1112.0860}{arXiv:1112.0860 [hep-ph]}\BibitemShut
  {NoStop}%
%%CITATION = ARXIV:1112.0860;%%
\bibitem [{\citenamefont {Manohar}\ and\ \citenamefont
  {Georgi}(1984)}]{Manohar:1983md}%
  \BibitemOpen
  \bibfield  {author} {\bibinfo {author} {\bibfnamefont {A.}~\bibnamefont
  {Manohar}}\ and\ \bibinfo {author} {\bibfnamefont {H.}~\bibnamefont
  {Georgi}},\ }\href {\doibase 10.1016/0550-3213(84)90231-1} {\bibfield
  {journal} {\bibinfo  {journal} {Nucl. Phys.}\ }\textbf {\bibinfo {volume}
  {B234}},\ \bibinfo {pages} {189} (\bibinfo {year} {1984})}\BibitemShut
  {NoStop}%
%%CITATION = NUPHA,B234,189;%%
\bibitem [{\citenamefont {Wong}(1998)}]{Wong:1998ex}%
  \BibitemOpen
  \bibfield  {author} {\bibinfo {author} {\bibfnamefont {S.~S.~M.}\
  \bibnamefont {Wong}},\ }\href@noop {} {\emph {\bibinfo {title} {{Introductory
  nuclear physics}}}}\ (\bibinfo {year} {1998})\BibitemShut {NoStop}%
%%CITATION = INSPIRE-484484;%%
\bibitem [{\citenamefont {Martin}(1997)}]{Martin:1997ns}%
  \BibitemOpen
  \bibfield  {author} {\bibinfo {author} {\bibfnamefont {S.~P.}\ \bibnamefont
  {Martin}},\ }\href {\doibase 10.1142/9789812839657_0001,
  10.1142/9789814307505_0001} {\ ,\ \bibinfo {pages} {1} (\bibinfo {year}
  {1997})},\ \bibinfo {note} {[Adv. Ser. Direct. High Energy
  Phys.18,1(1998)]},\ \Eprint
  {http://arxiv.org/abs/hep-ph/9709356}{arXiv:hep-ph/9709356
  [hep-ph]}\BibitemShut {NoStop}%
%%CITATION = HEP-PH/9709356;%%
\bibitem [{\citenamefont {Cirelli}\ \emph {et~al.}(2006)\citenamefont
  {Cirelli}, \citenamefont {Fornengo},\ and\ \citenamefont
  {Strumia}}]{Cirelli:2005uq}%
  \BibitemOpen
  \bibfield  {author} {\bibinfo {author} {\bibfnamefont {M.}~\bibnamefont
  {Cirelli}}, \bibinfo {author} {\bibfnamefont {N.}~\bibnamefont {Fornengo}}, \
  and\ \bibinfo {author} {\bibfnamefont {A.}~\bibnamefont {Strumia}},\ }\href
  {\doibase 10.1016/j.nuclphysb.2006.07.012} {\bibfield  {journal} {\bibinfo
  {journal} {Nucl. Phys.}\ }\textbf {\bibinfo {volume} {B753}},\ \bibinfo
  {pages} {178} (\bibinfo {year} {2006})},\ \Eprint
  {http://arxiv.org/abs/hep-ph/0512090}{arXiv:hep-ph/0512090
  [hep-ph]}\BibitemShut {NoStop}%
%%CITATION = HEP-PH/0512090;%%
\bibitem [{\citenamefont {Toharia}\ and\ \citenamefont
  {Wells}(2006)}]{Toharia:2005gm}%
  \BibitemOpen
  \bibfield  {author} {\bibinfo {author} {\bibfnamefont {M.}~\bibnamefont
  {Toharia}}\ and\ \bibinfo {author} {\bibfnamefont {J.~D.}\ \bibnamefont
  {Wells}},\ }\href {\doibase 10.1088/1126-6708/2006/02/015} {\bibfield
  {journal} {\bibinfo  {journal} {JHEP}\ }\textbf {\bibinfo {volume} {02}},\
  \bibinfo {pages} {015} (\bibinfo {year} {2006})},\ \Eprint
  {http://arxiv.org/abs/hep-ph/0503175}{arXiv:hep-ph/0503175
  [hep-ph]}\BibitemShut {NoStop}%
%%CITATION = HEP-PH/0503175;%%
\bibitem [{\citenamefont {Gambino}\ \emph {et~al.}(2005)\citenamefont
  {Gambino}, \citenamefont {Giudice},\ and\ \citenamefont
  {Slavich}}]{Gambino:2005eh}%
  \BibitemOpen
  \bibfield  {author} {\bibinfo {author} {\bibfnamefont {P.}~\bibnamefont
  {Gambino}}, \bibinfo {author} {\bibfnamefont {G.~F.}\ \bibnamefont
  {Giudice}}, \ and\ \bibinfo {author} {\bibfnamefont {P.}~\bibnamefont
  {Slavich}},\ }\href {\doibase 10.1016/j.nuclphysb.2005.08.011} {\bibfield
  {journal} {\bibinfo  {journal} {Nucl. Phys.}\ }\textbf {\bibinfo {volume}
  {B726}},\ \bibinfo {pages} {35} (\bibinfo {year} {2005})},\ \Eprint
  {http://arxiv.org/abs/hep-ph/0506214}{arXiv:hep-ph/0506214
  [hep-ph]}\BibitemShut {NoStop}%
%%CITATION = HEP-PH/0506214;%%
\bibitem [{\citenamefont {Sato}\ \emph {et~al.}(2012)\citenamefont {Sato},
  \citenamefont {Shirai},\ and\ \citenamefont {Tobioka}}]{Sato:2012xf}%
  \BibitemOpen
  \bibfield  {author} {\bibinfo {author} {\bibfnamefont {R.}~\bibnamefont
  {Sato}}, \bibinfo {author} {\bibfnamefont {S.}~\bibnamefont {Shirai}}, \ and\
  \bibinfo {author} {\bibfnamefont {K.}~\bibnamefont {Tobioka}},\ }\href
  {\doibase 10.1007/JHEP11(2012)041} {\bibfield  {journal} {\bibinfo  {journal}
  {JHEP}\ }\textbf {\bibinfo {volume} {11}},\ \bibinfo {pages} {041} (\bibinfo
  {year} {2012})},\ \Eprint {http://arxiv.org/abs/1207.3608}{arXiv:1207.3608
  [hep-ph]}\BibitemShut {NoStop}%
%%CITATION = ARXIV:1207.3608;%%
\bibitem [{\citenamefont {Sato}\ \emph {et~al.}(2013)\citenamefont {Sato},
  \citenamefont {Shirai},\ and\ \citenamefont {Tobioka}}]{Sato:2013bta}%
  \BibitemOpen
  \bibfield  {author} {\bibinfo {author} {\bibfnamefont {R.}~\bibnamefont
  {Sato}}, \bibinfo {author} {\bibfnamefont {S.}~\bibnamefont {Shirai}}, \ and\
  \bibinfo {author} {\bibfnamefont {K.}~\bibnamefont {Tobioka}},\ }\href
  {\doibase 10.1007/JHEP10(2013)157} {\bibfield  {journal} {\bibinfo  {journal}
  {JHEP}\ }\textbf {\bibinfo {volume} {10}},\ \bibinfo {pages} {157} (\bibinfo
  {year} {2013})},\ \Eprint {http://arxiv.org/abs/1307.7144}{arXiv:1307.7144
  [hep-ph]}\BibitemShut {NoStop}%
%%CITATION = ARXIV:1307.7144;%%
\bibitem [{\citenamefont {Fukuda}\ \emph {et~al.}()\citenamefont {Fukuda},
  \citenamefont {Luo},\ and\ \citenamefont {Shirai}}]{gluinoInprep}%
  \BibitemOpen
  \bibfield  {author} {\bibinfo {author} {\bibfnamefont {H.}~\bibnamefont
  {Fukuda}}, \bibinfo {author} {\bibfnamefont {F.}~\bibnamefont {Luo}}, \ and\
  \bibinfo {author} {\bibfnamefont {S.}~\bibnamefont {Shirai}},\ }\href@noop {}
  {\bibinfo  {journal} {in preparation}\ }\BibitemShut {NoStop}%
\bibitem [{\citenamefont {Saikawa}\ and\ \citenamefont
  {Shirai}(2018)}]{Saikawa:2018rcs}%
  \BibitemOpen
\bibfield  {journal} {  }\bibfield  {author} {\bibinfo {author} {\bibfnamefont
  {K.}~\bibnamefont {Saikawa}}\ and\ \bibinfo {author} {\bibfnamefont
  {S.}~\bibnamefont {Shirai}},\ }\href {\doibase 10.1088/1475-7516/2018/05/035}
  {\bibfield  {journal} {\bibinfo  {journal} {JCAP}\ }\textbf {\bibinfo
  {volume} {1805}},\ \bibinfo {pages} {035} (\bibinfo {year} {2018})},\ \Eprint
  {http://arxiv.org/abs/1803.01038}{arXiv:1803.01038 [hep-ph]}\BibitemShut
  {NoStop}%
%%CITATION = ARXIV:1803.01038;%%
\bibitem [{\citenamefont {Aghanim}\ \emph {et~al.}(2018)\citenamefont {Aghanim}
  \emph {et~al.}}]{Aghanim:2018eyx}%
  \BibitemOpen
  \bibfield  {author} {\bibinfo {author} {\bibfnamefont {N.}~\bibnamefont
  {Aghanim}} \emph {et~al.} (\bibinfo {collaboration} {Planck}),\ }\href@noop
  {} {\  (\bibinfo {year} {2018})},\ \Eprint
  {http://arxiv.org/abs/1807.06209}{arXiv:1807.06209 [astro-ph.CO]}\BibitemShut
  {NoStop}%
%%CITATION = ARXIV:1807.06209;%%
\bibitem [{\citenamefont {Aprile}\ \emph {et~al.}(2018)\citenamefont {Aprile}
  \emph {et~al.}}]{Aprile:2018dbl}%
  \BibitemOpen
  \bibfield  {author} {\bibinfo {author} {\bibfnamefont {E.}~\bibnamefont
  {Aprile}} \emph {et~al.} (\bibinfo {collaboration} {XENON}),\ }\href
  {\doibase 10.1103/PhysRevLett.121.111302} {\bibfield  {journal} {\bibinfo
  {journal} {Phys. Rev. Lett.}\ }\textbf {\bibinfo {volume} {121}},\ \bibinfo
  {pages} {111302} (\bibinfo {year} {2018})},\ \Eprint
  {http://arxiv.org/abs/1805.12562}{arXiv:1805.12562 [astro-ph.CO]}\BibitemShut
  {NoStop}%
%%CITATION = ARXIV:1805.12562;%%
\bibitem [{\citenamefont {Drees}\ and\ \citenamefont
  {Nojiri}(1993)}]{Drees:1993bu}%
  \BibitemOpen
  \bibfield  {author} {\bibinfo {author} {\bibfnamefont {M.}~\bibnamefont
  {Drees}}\ and\ \bibinfo {author} {\bibfnamefont {M.}~\bibnamefont {Nojiri}},\
  }\href {\doibase 10.1103/PhysRevD.48.3483} {\bibfield  {journal} {\bibinfo
  {journal} {Phys. Rev.}\ }\textbf {\bibinfo {volume} {D48}},\ \bibinfo {pages}
  {3483} (\bibinfo {year} {1993})},\ \Eprint
  {http://arxiv.org/abs/hep-ph/9307208}{arXiv:hep-ph/9307208
  [hep-ph]}\BibitemShut {NoStop}%
%%CITATION = HEP-PH/9307208;%%
\bibitem [{\citenamefont {Gondolo}\ and\ \citenamefont
  {Scopel}(2013)}]{Gondolo:2013wwa}%
  \BibitemOpen
  \bibfield  {author} {\bibinfo {author} {\bibfnamefont {P.}~\bibnamefont
  {Gondolo}}\ and\ \bibinfo {author} {\bibfnamefont {S.}~\bibnamefont
  {Scopel}},\ }\href {\doibase 10.1088/1475-7516/2013/10/032} {\bibfield
  {journal} {\bibinfo  {journal} {JCAP}\ }\textbf {\bibinfo {volume} {1310}},\
  \bibinfo {pages} {032} (\bibinfo {year} {2013})},\ \Eprint
  {http://arxiv.org/abs/1307.4481}{arXiv:1307.4481 [hep-ph]}\BibitemShut
  {NoStop}%
%%CITATION = ARXIV:1307.4481;%%
\bibitem [{\citenamefont {Hisano}\ \emph {et~al.}(2011)\citenamefont {Hisano},
  \citenamefont {Ishiwata},\ and\ \citenamefont {Nagata}}]{Hisano:2011um}%
  \BibitemOpen
  \bibfield  {author} {\bibinfo {author} {\bibfnamefont {J.}~\bibnamefont
  {Hisano}}, \bibinfo {author} {\bibfnamefont {K.}~\bibnamefont {Ishiwata}}, \
  and\ \bibinfo {author} {\bibfnamefont {N.}~\bibnamefont {Nagata}},\ }\href
  {\doibase 10.1016/j.physletb.2011.11.017} {\bibfield  {journal} {\bibinfo
  {journal} {Phys. Lett.}\ }\textbf {\bibinfo {volume} {B706}},\ \bibinfo
  {pages} {208} (\bibinfo {year} {2011})},\ \Eprint
  {http://arxiv.org/abs/1110.3719}{arXiv:1110.3719 [hep-ph]}\BibitemShut
  {NoStop}%
%%CITATION = ARXIV:1110.3719;%%
\bibitem [{\citenamefont {Abdel-Rehim}\ \emph {et~al.}(2016)\citenamefont
  {Abdel-Rehim}, \citenamefont {Alexandrou}, \citenamefont {Constantinou},
  \citenamefont {Hadjiyiannakou}, \citenamefont {Jansen}, \citenamefont
  {Kallidonis}, \citenamefont {Koutsou},\ and\ \citenamefont {Vaquero
  Aviles-Casco}}]{Abdel-Rehim:2016won}%
  \BibitemOpen
  \bibfield  {author} {\bibinfo {author} {\bibfnamefont {A.}~\bibnamefont
  {Abdel-Rehim}}, \bibinfo {author} {\bibfnamefont {C.}~\bibnamefont
  {Alexandrou}}, \bibinfo {author} {\bibfnamefont {M.}~\bibnamefont
  {Constantinou}}, \bibinfo {author} {\bibfnamefont {K.}~\bibnamefont
  {Hadjiyiannakou}}, \bibinfo {author} {\bibfnamefont {K.}~\bibnamefont
  {Jansen}}, \bibinfo {author} {\bibfnamefont {C.}~\bibnamefont {Kallidonis}},
  \bibinfo {author} {\bibfnamefont {G.}~\bibnamefont {Koutsou}}, \ and\
  \bibinfo {author} {\bibfnamefont {A.}~\bibnamefont {Vaquero Aviles-Casco}}
  (\bibinfo {collaboration} {ETM}),\ }\href {\doibase
  10.1103/PhysRevLett.116.252001} {\bibfield  {journal} {\bibinfo  {journal}
  {Phys. Rev. Lett.}\ }\textbf {\bibinfo {volume} {116}},\ \bibinfo {pages}
  {252001} (\bibinfo {year} {2016})},\ \Eprint
  {http://arxiv.org/abs/1601.01624}{arXiv:1601.01624 [hep-lat]}\BibitemShut
  {NoStop}%
%%CITATION = ARXIV:1601.01624;%%
\bibitem [{\citenamefont {Kovarik}\ \emph {et~al.}(2016)\citenamefont {Kovarik}
  \emph {et~al.}}]{Kovarik:2015cma}%
  \BibitemOpen
  \bibfield  {author} {\bibinfo {author} {\bibfnamefont {K.}~\bibnamefont
  {Kovarik}} \emph {et~al.},\ }\href {\doibase 10.1103/PhysRevD.93.085037}
  {\bibfield  {journal} {\bibinfo  {journal} {Phys. Rev.}\ }\textbf {\bibinfo
  {volume} {D93}},\ \bibinfo {pages} {085037} (\bibinfo {year} {2016})},\
  \Eprint {http://arxiv.org/abs/1509.00792}{arXiv:1509.00792
  [hep-ph]}\BibitemShut {NoStop}%
%%CITATION = ARXIV:1509.00792;%%
\bibitem [{\citenamefont {Gross}\ \emph {et~al.}(2018)\citenamefont {Gross},
  \citenamefont {Mitridate}, \citenamefont {Redi}, \citenamefont {Smirnov},\
  and\ \citenamefont {Strumia}}]{Gross:2018zha}%
  \BibitemOpen
  \bibfield  {author} {\bibinfo {author} {\bibfnamefont {C.}~\bibnamefont
  {Gross}}, \bibinfo {author} {\bibfnamefont {A.}~\bibnamefont {Mitridate}},
  \bibinfo {author} {\bibfnamefont {M.}~\bibnamefont {Redi}}, \bibinfo {author}
  {\bibfnamefont {J.}~\bibnamefont {Smirnov}}, \ and\ \bibinfo {author}
  {\bibfnamefont {A.}~\bibnamefont {Strumia}},\ }\href@noop {} {\  (\bibinfo
  {year} {2018})},\ \Eprint {http://arxiv.org/abs/1811.08418}{arXiv:1811.08418
  [hep-ph]}\BibitemShut {NoStop}%
%%CITATION = ARXIV:1811.08418;%%
\end{thebibliography}%
%%%%%%%%%%%%%%%%%%%%%%%%%%%%%%
%%%%%%%%%%%%%%

\end{document}